# The Distance of GRB is Independent from the Redshift

Fu-Gao Song [1,2]

[1] College of Electronic Science and Technology, Shenzhen University, Shenzhen, 518060, China
[2] Shenzhen Key Laboratory of Micro-nano Photonic Information Technology, Shenzhen, China

**Abstract.** Although it is acknowledged that GRB redshift is cosmological at present, but it in fact has never been confirmed. However, the fact is still unclear because which may be cosmological redshift (including possibly the host galaxy redshift or the background galaxy redshift) or the redshift results from gravity of neutron star; in other words, GRB redshift may be related to its distance or not. Here, I enumerate a series of evidences, including three methods, to determine whether the GRB distance does depend on the redshift. Firstly, the correlation analysis shows that there is no correlation between the fluence of 131 GRBs (and the 1 s peak photon flux of 111 GRBs) and the redshift although there is a significant correlation between the apparent magnitude of 32 hosts and the redshift. Secondly, from the number-redshift relation of GRBs and the deductive reasoning, one can deduce an absurd conclusion that the probability of a nearby galaxy generating a GRB event would vary inversely as its distance square if GRB indeed comes from an external galaxy and the distance depends on the redshift. Thirdly, if the distance is related to the redshift, then the variables of fluence and peak flux definitely cannot be separated from the variable of redshift in distribution functions of both the fluence-redshift and the peak flux-redshift; while the variables separation tests show that they in effect do, and we then can exactly forecast the values of the fluence and the peak flux for the GRBs with redshift $z > 4.5$. Other evidences all show that GRB distance is independent from the redshift without exception.

## 1. INTRODUCTION

The event of gamma-ray burst (GRB) was discovered in the late 1960's; after several years, the report was published (Klebesadel et al. 1973); due to the difficulty on observation, their nature and origin remained thereafter a mystery for more than three decades.

The launch of the Compton Gamma-Ray Observatory (CGRO) in 1991 opened a new era for the research in GRB; over 2700 bursts had been recorded by the all-sky survey of the Burst and Transient Source Experiment (BATSE) on CGRO, which showed that the angular distribution of GRBs is generally isotropic (Meegen et al. 1992; Briggs et al. 1996), this implies that either the bursts are located in an extended galactic halo (Paczyński 1991) or their origin is cosmological (van den Bergh 1983; Paczyński 1986).

A decisive change for the research of GRB occurred in 1997 when the Italian-Dutch satellite Beppo-SAX launches with success; which result in the discovery of the redshift for a number of GRBs (Metzger et al. 1997; Odewahn et al. 1998; Tinney et al. 1998; Djorgovski et al. 1999a, 1999b, 1999c; Vreeswijk et al. 1999a, 1999b; see also Table 1 below). In fact, in order to seek the evidences for the cosmological origin of GRB, the redshift is long-awaited; therefore, its appearing was thought to be an irrefutable evidence for the cosmological origin of GRB at once. Currently, it is widely accepted that GRBs are a transitory physical process with extreme high energy occurring in a faraway host galaxy and the distance is completely determined by the redshift, which was also called as the host redshift although the host even in effect does not been detected. This is the actual status quo! However, the conclusion that the distance depends on the redshift in fact has never been confirmed formally up to now.

Does the GRB distance indeed depend on the redshift? This is a significant topic, and is worth to further confirm. In fact, there are a series of evidences to show that the GRB distance does not depend on the redshift; if we force the redshift to indicate the distance, we would inevitably obtain many wrong conclusions that violates the basic principle of physics.

On the other hand, some problems regarding GRB redshift were recently discussed by D. Coward (Coward 2007). This discussion shows that there are problems there; there are also a number of similar discussions in the literature (Daigne et al. 2006; Truong & Dermer 2006; Firmani et al. 2004; Fruchter et al. 2006; Stanek et al. 2006; Fiore et al. 2006). However, the essential of problems is in whether the GRB distance does depend on the redshift instead of in the others; the problems would not exist if the distance does not depend on the redshift.

Although it is difficult to determine if the distance of a single GRB does depend on the redshift, however, the problem can be reliably solved by many methods for a great number of GRBs, see below.

In the last decade, the redshift of a great lot of GRBs had been determined.

Among those GRBs, some other physical quantities have also been determined, such as the fluence, the 1 s peak energy flux, the 1 s peak photon flux, the flux of the X-ray transients (XT), the flux of the radio transients (RT), the apparent magnitude of the optical transients (OT) and that of partial hosts themselves and others do; among these physical quantities, except for partial redshift and the apparent magnitude of hosts are inherent in the host and the apparent magnitude depends on its distance, the others are all inherent in GRBs, their magnitude only depends on the distance of GRBs. If GRBs distance really depends on the redshift, then the distribution states of all apparent physical quantities of GRBs must be related to the redshift and which can be easily identified; we can therefore confirm whether GRBs were generated in the hosts.

Table 1 collected 154 GRBs that redshift has been determined before 1 Jan 2008 (up to GRB071227), but only the first 132 GRBs (up to GRB070318) is used in the paper due to the historical causes; among them, the fluence of 131 GRBs and the 1 s peak photon flux of 111 GRBs have also been determined, also the apparent magnitude of 32 hosts does (see also Table 1 below). Due to the apparent brightness of XT, RT and OT vary rapidly with the time and the observed data are therefore not standard, these data are not used in the analysis. Using the four kinds of data above, we can definitely prove that GRB distance is independent from the redshift and the GRB redshift is arising from either the background galaxy or the gravity.

**Table 1 Summary of the observed data of GRBs**

(The full data see below)

## 2. KNOTTINESSES FROM LUMINOSITY DISTANCE

Note that there is no discrepancy between both kinds of GRBs detected by BATSE and that determined redshift if remove the knowledge obtained from the afterglow observation. Therefore, in order to confirm if the distance depends on the redshift, we have sufficient reasons to compare both luminosity distances determined by redshift and by luminosity function. Moreover, the comparison analysis below also has solid statistic grounds because there are three statistics rules here:

**Rule 1.** The discrete degree of a physical quantity depends on the sample size. For example, let $f(x)$ be the normalized distribution function of the physical quantity $x > 0$, $N$ be the sample size; define the minimum $x_1$- and the maximum $x_2$ of the physical quantity as follows:

$$\int_0^\infty f(x)dx = 1, \quad N\int_0^{x_1} f(x)dx = 1 \quad \text{and} \quad N\int_{x_2}^\infty f(x)dx = 1;$$

we then know that the larger the $N$ is, the smaller the minimum $x_1$ and the greater the maximum $x_2$ will be. If define the degree of dispersion of physical quantity $x$ as $x_2/x_1$, then there must be that the value of $x_2/x_1$ for large sample is greater than that for small sample. Therefore, the results of the comparison analysis about the degree of dispersion below are completely reliable because the number of GRBs detected by BATSE is much greater than that of GRBs with determined redshift.

**Rule 2.** The physical quantity values of the most of samples would distribute around the most probable value $x_0$; e.g., the normal distribution gives

$$f(x) = (2\pi)^{-1/2} \exp(-x^2/2) \text{ with } \int_{-\infty}^{\infty} f(x)dx = 1, \text{ and}$$

$$\int_{-1}^{1} f(x)dx = 0.68273, \quad \int_{-2}^{2} f(x)dx = 0.95451;$$

i.e. 68.3% of samples would take the value within $(x_0-\sigma, x_0+\sigma)$; 95.4% of samples within $(x_0-2\sigma, x_0+2\sigma)$. In other words, the standard candle analysis and the correlation analysis have their solid statistic grounds.

**Rule 3.** An event with small probability can be neglected in finite trials (or small sample), this is an acknowledged rule. Therefore, if an event has already occurred in finite trials, and if its probability consists of a product of two probabilities and one of them is known to be small, then another must close to its maximum such that the event is not a small probability event.

The above three rules are the analysis foundations of this paper.

Let's simply compare the degree of dispersion for both the luminosity distances obtained from the statistics method and the redshift, respectively, below. We put the deceleration parameter to be $q_0 = 0.2$ in the paper.

Let $f$ express the fluence, $p$ the peak photon flux; define the luminosities $F$ and $P$ as follows:

$$F = 4\pi d_f^2 f, \quad P = 4\pi d_p^2 p. \tag{2.1}$$

Where, $d_f = d_p = d$ is the luminosity distances of the fluence and the peak photon flux:

$$d = \frac{c\left(q_0 z + (q_0 - 1)[(1 + 2q_0 z)^{1/2} - 1]\right)}{H_0 q_0^2 (1+z)^{1/2}} = \frac{c}{H_0} \phi_1(q_0, z), \tag{2.2}$$

$H_0$ is the Hubble constant, $c$ the speed of light.

There are 2132 GRBs detected by BATSE there, which the peak photon fluxes with 64 ms, 256 ms and 1024 ms timescales and total fluence were determined; their true distance does not be determined, but their relative distance can be estimated by statistic method. It is well known that the farther the object is, the fainter it will be; the closer the object is, the brighter it will be. Therefore, the distance of GRBs can be estimated by means of the brightness. From equation (2.1), the estimations of the relative distances for the faintest GRB to the brightest one can be obtained as follows:

$$D_p = \frac{d_{p,\text{faintest}}}{d_{p,\text{brightest}}} = \sqrt{\frac{P_{\text{faintest}}}{P_{\text{brightest}}}} \sqrt{\frac{p_{\text{brightest}}}{p_{\text{faintest}}}}, \tag{2.3}$$

$$D_f = \frac{d_{f,\text{faintest}}}{d_{f,\text{brightest}}} = \sqrt{\frac{F_{\text{faintest}}}{F_{\text{brightest}}}} \sqrt{\frac{f_{\text{brightest}}}{f_{\text{faintest}}}}. \qquad (2.4)$$

The minimum- and the maximum values for the peak photon flux of 64 ms, 256 ms and 1024 ms and the fluence among 2132 GRBs are found as follows respectively:

(*a*) 0.315 ph cm$^{-2}$ s$^{-1}$ (GRB910629B) and 183.370 ph cm$^{-2}$ s$^{-1}$ (GRB960924);

(*b*) 0.271 ph cm$^{-2}$ s$^{-1}$ (GRB950114) and 181.634 ph cm$^{-2}$ s$^{-1}$ (GRB960924);

(*c*) 0.051 ph cm$^{-2}$ s$^{-1}$ (GRB930406C) and 163.344 ph cm$^{-2}$ s$^{-1}$ (GRB960924);

(*d*) 1.591×10$^{-8}$ erg cm$^{-2}$ (GRB930203C) and 7.807×10$^{-4}$ erg cm$^{-2}$ (GRB940703).

It is clear that above four pairs of GRBs are not all the closest and farthest pair; in which there are at least three GRB pairs which the minimum values and the maximum values are caused by the diffusivity of the luminosities. However, in terms of the Rule 3 above, we can determine which one of them is the optimum closest- and farthest GRB pair. The deducing process is as follows:

1. GRBs are distributed in three dimensions space, the probability that the distance of a GRB occurs in the distance shell $x \sim x + \mathrm{d}x$ is proportional to $x^2$ for fixed $\mathrm{d}x$; therefore, the spatial distribution probability for the closest GRB (with the smallest $x$) is hence small, and then its (fluxes and fluence) luminosities must close to the most probable value.

2. Due to the finite instrument sensitivity, the farther GRBs are too faint to detect; therefore, the luminosity of a GRB detected as the farthest one would be possible higher than the most probable value (with larger probability) or close to it (with less probability).

3. The luminosity distance of a GRB should be selfsame no matter for three kinds of flux luminosities or for fluence luminosity; let $D_{p,64}$, $D_{p,256}$, $D_{p,1024}$ and $D_f$ denote the relative luminosity distance of three kinds of peak photon fluxes and the total fluence for two GRBs, respectively; there must be

$$D_{p,64} = D_{p,256} = D_{p,1024} = D_f. \qquad (2.5)$$

4. A GRB luminosity takes the real most probable value is difficult; however, the probability that four kinds of luminosity close to the most probable values simultaneously is the maximum, in which we have the following results:

$$D_{p,64} \cong D_{p,256} \cong D_{p,1024} \cong D_f. \qquad (2.6)$$

Now we calculate the four kinds of luminosity distance for above four GRB pairs, the results obtained from (*a*) to (*d*) are as follows:

(*a*) $D_{p,64} = 24.13$, $D_{p,256} = 24.24$, $D_{p,1024} = 25.87$ and $D_f = 26.03$ for the 1st pair;

(*b*) $D_{p,64} = 21.20$, $D_{p,256} = 25.89$, $D_{p,1024} = 26.82$ and $D_f = 6.61$ for the 2nd pair;

(*c*) $D_{p,64} = 14.35$, $D_{p,256} = 25.56$, $D_{p,1024} = 56.59$ and $D_f = 40.08$ for the 3rd pair;

(*d*) $D_{p,64} = 6.53$, $D_{p,256} = 8.57$, $D_{p,1024} = 13.53$ and $D_f = 221.55$ for the 4th pair.

We are sure that, from above results, the first GRB pair is precisely the optimum closest and farthest GRB pair, i.e. GRB960924 is the closest GRB (which is selected by three pairs) and GRB910629B is the farthest one. The mean value of the maximum relative distance is $25.076 \pm 0.949$, or $25.076 \times (1 \pm 0.038)$, the error is only $\pm 3.8\%$.

Now, we discuss the results obtained from the redshift distance. The minimum redshift in Table 1 is 0.0085 (GRB980425), the maximum is 6.6 (GRB060116). Although people think that GRB980425 is a particular GRB, but the comments are all based upon the afterglow observation; GRB980425 has in fact no discrepancy from those GRBs detected by BATSE if the information obtained from the afterglow observation is removed; we are interested only in does the distance depend on the redshift and no one has showed that the distance of GRB980425 is independent from its redshift up to now. Therefore, according to equation (2.2), the maximum of relative luminosity distance of GRBs would be 711 if the distance indeed depends on the redshift, and which is 28 times as much as that obtained from the brightness comparison. What does the numeral of 711 imply? Is it a reasonable value? In order to answer the problems, let's assume that the luminosity of the 64 ms peak photon flux of 2132 GRBs detected by BATSE are all selfsame and the maximum of their relative distance is precisely 711, from equation (2.1) we obtain that the maximum of the 64 ms relative peak photon flux would be $p_{max} / p_{min} = 711^2 = 5.06 \times 10^5$. While, as mentioned above, the first GRB pair gives the actual value is 582, which is only 1/869 times of the former! In other words, the redshift distance is entirely incompatible with the observed data of large sample. This first knottiness is caused by the redshift distance.

In addition, from equation (2.1) we obtain the relative luminosity of two GRBs as follows:

$$\frac{P}{P_0} = \left(\frac{d}{d_0}\right)^2 \frac{p}{p_0}. \qquad (2.7)$$

Table 1 given that $z = 0.0085$ and $p = 0.96$ ph cm$^{-2}$ s$^{-1}$ for GRB980425, $z = 0.16854$ and $p = 451$ ph cm$^{-2}$ s$^{-1}$ for GRB030329, and $z = 3.198$ and $p = 64$ ph cm$^{-2}$ s$^{-1}$ for GRB020124; using these data and equations (2.7) and (2.2) we obtain that the relative luminosity for GRB030329 to GRB980425 is $P / P_0 = 1.81 \times 10^5$ and that for GRB020124 to GRB980425 is $P / P_0 = 8.56 \times 10^6$, respectively; in other words, the distribution range of the relative luminosity of GRBs would at least reach to $1.81 \times 10^5$, even to $8.56 \times 10^6$. Are these values reasonable? In order to answer the problem, let's put 2132 GRBs detected by BATSE at a selfsame distance and assume their 1 s peak photon flux relative luminosity distribution range to be above values, then the maximum of the observed value of relative 1 s peak photon flux would at least reach to $1.81 \times 10^5$ even to $8.56 \times 10^6$. While, the third GRB pair above gives the actual maximum of 2132 GRBs is only 3203, which is only 1 / 57 and 1 / 2672 times of those above, respectively! Due to the lower redshift, the cosmological ages of GRB980425 and GRB030329 are almost the same, the evolutionary effects of the luminosity should be neglected; therefore, the huge discrepancy between the theoretical expectation and the actual value cannot be solved by the luminosity evolution. This is the second knottiness caused by the redshift distance.

The above facts that the theoretical lower bounds of physical quantities deducing from the redshift distance are much greater than the observed values are undoubtedly a deadly blow for the redshift distance. We cannot imagine what physics mechanism can solve the knottiness. In above estimations, the observed values are no problem, only the redshift distance is questionable; the problem is precisely caused by it. This shows clearly that the redshift cannot indicate the distance for GRBs.

### 3. CORRELATION ANALYSIS FOR GRBS, QUASARS AND HOSTS

Now let's use the standard statistics method to analyze the observed data of GRBs below. From equation (2.1) we obtain

$$\left.\begin{array}{l}\log f = \log(F/4\pi) - 2\log d_f \\ \log p = \log(P/4\pi) - 2\log d_p\end{array}\right\}. \qquad (3.1)$$

We are then sure that there must exist correlation between $\log f$ and $\log p$ no matter whether there is correlation or not between $F$ and $P$.

FIG 1(a) showed the correlation between $\log f$ and $\log p$ for 110 GRBs from Table 1, which given the correlation coefficient to be $r = 0.600$; FIG. 1(b) showed that for 2132 GRBs detected by BATSE with $r = 0.765$. These show that the correlation between $\log f$ and $\log p$ of GRBs is very significant.

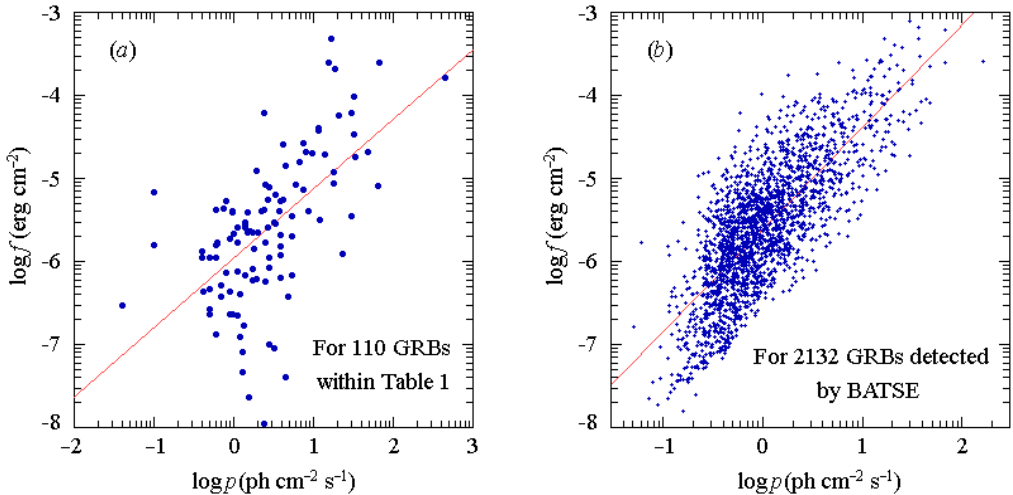

**FIG. 1.** The $\log f$-$\log p$ relations of GRBs; where the lines express the regression results. (*a*) The $\log f$-$\log p$ relation for 110 GRBs cited from Table 1. (*b*) The log $f$-$\log p$ relation for 2132 GRBs detected by BATSE.

The statistic correlation between apparent magnitude and redshift can be emerged by two methods: the statistics method and a graphic method; each of them has its virtue: the former is accurate and the latter is intuitional. Both methods are used in the paper.

In order to show the rationality of correlation analysis for confirming whether the distances of both GRBs and hosts are selfsame, we firstly introduce some examples of correlation analysis.

Due to quasars and host galaxies are located in the depth of the Universe, there must exist the statistical correlation between the apparent magnitude and the distance (redshift) of them.

The brightness of a cosmic object would depend on its distance (or redshift). In Friedman's standard cosmology, the relation between the apparent magnitude $m$ of an object and the redshift $z$ was shown as follows:

$$m = m_0 + 5\log\phi(q_0, z). \qquad (3.2)$$

Where, $m_0$ is a constant, and

$$\phi(q_0, z) = q_0^{-2}\left(q_0 z + (q_0 - 1)[(1 + 2q_0 z)^{1/2} - 1]\right). \qquad (3.3)$$

The luminosity of different cosmic objects is generally not identical; sometimes, they would have a great difference. However, the luminosity of them in general would distribute by definite luminosity function that would have a maximum corresponding to the most probable luminosity of galaxies as mentioned above. It is infrequent that a luminosity of an object keeps away from the most probable value. Therefore, it can be expected that the luminosities of the most objects would close to the most probable value; in other words, the luminosities of hosts (or quasars) should be approximately equal to each other. Therefore, hosts and quasars should display the correlation as showing in equation (3.2).

In order to verify this conclusion, let's do the correlation analysis by following formula:

$$m = a + 5b\log\phi(q_0, z). \qquad (3.4)$$

It can be expected that the values of both the regression coefficient $b$ and the correlation coefficient $r$ would close to 1 in the result of correlation analysis for the hosts and the quasars.

Due to quasars indeed come from the depth of the Universe, according to above reasons, there must exist the significant correlation between their apparent magnitude and the redshift. The expectation can be confirmed. Indeed, there is a subset of QSO77292 containing 77292 quasars that both the redshift and the $U$-band magnitude have been determined in the catalogue of dr5qso.dat obtained from the Sloan Digital Sky Survey (see also http://www.sdss.org/dr5/products/value_added/qsocat_dr5.html);

the apparent magnitude-redshift distribution has been showed in FIG. 2(*a*); and the results of correlation analysis for different deceleration parameters $q_0$ were shown in the Table 2.

FIG. 2(*a*) showed that quasars distribute densely in a diagonal zone with three pairs parallel sides (the empiric structure of the diagonal zone see also FIG. 2), where contains 76953 quasars (99.56%); there are only 339 quasars in other place, in which 297 quasars are in the bottom left corner and only 42 quasars in the top right corner.

The distribution characteristics showing in FIG. 2(*a*) are clear enough: the quasars with the highest redshift are almost the faintest one; on the contrary, those with the lowest redshift are almost the brightest one. In other words, the higher the redshift, the fainter the quasars were; the lower the redshift, the brighter the quasars were. Those are precisely the expectation for both the theory and the general knowledge. The criterion of correlation for the graphic method is precisely in that the most of samples would almost distribute in the diagonal zone and there are practically negligible samples in both the top right corner and the bottom left corner; this embodies that the farther the object, the fainter it is; the closer the object, the brighter it is. If there are many samples (e.g. total over 10%) in both the top right corner and the bottom left corner, we can affirm, from the graph, that the correlation does not exist.

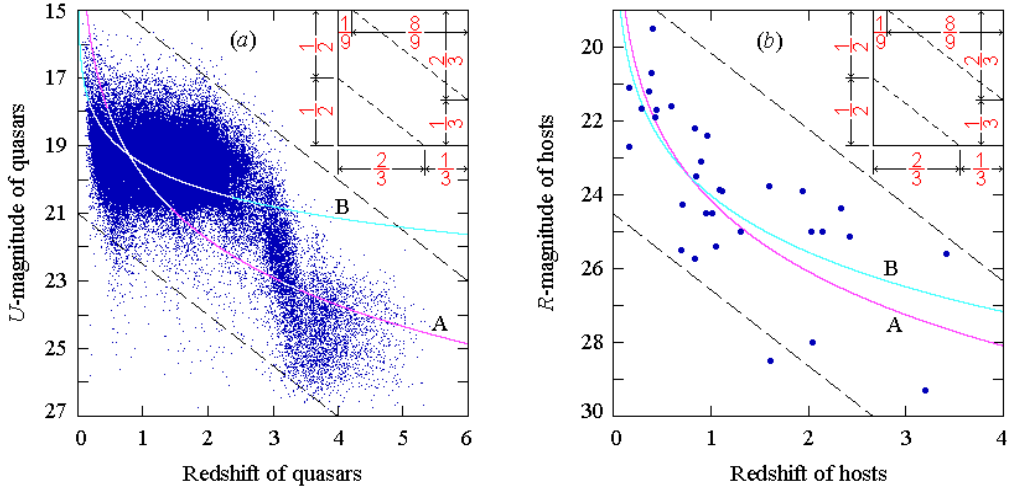

**FIG. 2.** Apparent magnitude-redshift relation of quasars and hosts. Where, curve A expresses the curve given by equation (3.2), curves B the regression curves, i.e. equation (3.3). The little figure on the top right corner in each figure is the empiric structure of the diagonal zone, which is used in the graphic method. (*a*) The *U-z* relation for the subset of QSO77292. (*b*) The *R-z* relation for the subset of Host32.

Table 2 shows that the values of both *b* and *r* reached 0.4 and 0.48 for quasars respectively, which imply that the correlation is significant in the case of quasars.

Only a few apparent magnitudes of hosts have been determined. In order to increase the sample size, we consider together those hosts that *R*-band or $R_c$-band, or *V*-band magnitude has been determined because those filters are relatively approach. There are thirty-two such hosts making up the subset of Host32 among the first 132 GRBs in Table 1. The apparent magnitude-redshift relation has been shown in FIG. 2(*b*), and the results of correlation analysis were also shown in the Table 2.

**Table 2. Summary of correlation analysis for different subsets**

| $q_0$ | | 0 | 0.1 | 0.2 | 0.3 | 0.4 | 0.5 |
|---|---|---|---|---|---|---|---|
| Subset of QSO77292 [1] | *r* | 0.5038 | 0.4929 | 0.4871 | 0.4834 | 0.4810 | 0.4793 |
| | *b* | 0.3903 | 0.4075 | 0.4224 | 0.4356 | 0.4473 | 0.4580 |
| | *a* | 19.164 | 19.221 | 19.270 | 19.315 | 19.356 | 19.394 |
| Subset of Host32 [2] | *r* | 0.7443 | 0.7434 | 0.7426 | 0.7418 | 0.7412 | 0.7406 |
| | *b* | 0.7440 | 0.7839 | 0.8154 | 0.8418 | 0.8648 | 0.8852 |
| | *a* | 23.321 | 23.420 | 23.510 | 23.594 | 23.673 | 23.747 |
| Subset of GRB131 [3] | *r* | −0.0508 | −0.0578 | −0.0619 | −0.0646 | −0.0665 | −0.0679 |
| | *b* | −0.0446 | −0.0545 | −0.0612 | −0.0663 | −0.0704 | −0.0738 |
| | *a* | −5.4897 | −5.4864 | −5.4826 | −5.4789 | −5.4752 | −5.4717 |
| Subset of GRB111 [4] | *r* | 0.0597 | 0.0492 | 0.0426 | 0.0377 | 0.0339 | 0.0308 |
| | *b* | 0.0349 | 0.0308 | 0.0280 | 0.0257 | 0.0239 | 0.0222 |
| | *a* | 0.4302 | 0.4262 | 0.4237 | 0.4220 | 0.4208 | 0.4199 |
| Subset of HF31 [5] | *r* | −0.0292 | −0.0363 | −0.0409 | −0.0443 | −0.0468 | −0.0487 |
| | *b* | −0.0315 | −0.0418 | −0.0495 | −0.0557 | -0.0609 | −0.0653 |
| | *a* | −4.8428 | −4.8399 | −4.8367 | −4.8333 | −4.8300 | −4.8267 |

[1] QSO77292 contain 77292 quasars that redshift and *U*-band magnitude have been determined in http://www.sdss.org/dr5/products/value_added/qsocat_dr5.html.
[2] Host32 contains 32 hosts that *R*-band or $R_c$-band, or *V*-band magnitude has been determined among the first 132 GRBs in Table 1.
[3] GRB131 contains 131 GRBs that redshift and fluence have been determined among the first 132 GRBs in Table 1.
[4] GRB111 contains 111 GRBs that redshift and 1 s peak photon number flux have been determined among the first 132 GRBs in Table 1.
[5] HF31 is the intersection of the GRB131 and the Host32 and contains 31 GRBs.

The distribution of hosts also has many characteristics:

1, total thirty-one hosts distribute in the diagonal zone among 32 samples.

2, all the eight brightest hosts with $R < 22$ satisfy $z < 0.6$.

3, all the closest hosts with $z < 0.6$ are the brightest one.

4, all the faintest hosts with $R > 24$ are the farthest one satisfying $z > 1.6$.

Those are precisely the theoretical expectation and have been showed clearly in FIG. 2(*b*).

It is evident from Table 2 that there is a more significant correlation in the case of hosts than that of quasars because both values of *b* and *r* have reached 0.74 in the case of hosts.

Above facts show that there must exist statistical correlation between two different apparent physical quantities with identical distance, and which demonstrate to us that in order to determine whether the redshift of hosts can indicate the distance of GRBs, we only need to verify the statistical correlation between each apparent physical quantity of GRBs and the redshift of hosts. As a result we can determine the true status of hosts.

## 4. CORRELATION ANALYSIS FOR GRB TO REDSHIFT

If GRB indeed come from the host, then the distance of the GRB must be determined by the redshift of the host, and then there must exist a similar significant correlation between each apparent physical quantity of GRBs and the redshift of hosts as quasars and hosts do. On the contrary, if GRB does not come from the host, then the correlation is impossible. Therefore, it can be determined whether GRBs come from the hosts by verifying the correlation between each apparent physical quantity of GRBs and the redshift of hosts.

From equations (2.1) and (2.2) we obtain

$$\log f = \log\left(\frac{H_0^2 F}{4\pi c^2}\right) - \log \phi_1^2(q_0, z) ; \qquad (4.1)$$

$$\log p = \log\left(\frac{H_0^2 P}{4\pi c^2}\right) - \log \phi_1^2(q_0, z) . \qquad (4.2)$$

In order to verify the correlation between *f* (and *p*) and the redshift, formulae (4.1) and (4.2) should be rewritten as follows:

$$\log f = a - b \log \phi_1^2(q_0, z) ; \qquad (4.3)$$

$$\log p = a - b \log \phi_1^2(q_0, z) . \qquad (4.4)$$

If GRBs indeed come from the hosts, then the values of both *b* and *r* would close to 1 just as the hosts do. Nevertheless, if they are so small even their sign is wrong in the result of correlation analysis, we can then affirm that there must be no affiliation between GRBs and hosts.

There are two subsets of GRBs among the first 132 GRBs in Table 1. The subset of GRB131 contains 131 GRBs which fluence *f* was determined; and the subset of GRB111 contains 111 GRBs which 1 s peak photon flux *p* was determined. The distributions of both physical quantities vs. redshift were shown in FIG. 3(*a*) and FIG. 3(*b*) respectively; the results of the correlation analysis were also shown in Table 2.

There are twenty-five samples in the bottom left corner and three in the top right corner in FIG. 3(*a*), total 21.4% of samples do not distribute in the diagonal zone, which implies that the correlation does not exist at present. There are two significant characteristics in the FIG. 3(*a*):
1, the GRBs with faintest fluences are all having the lowest redshift instead of the highest; in other words, the faintest GRBs would be not the farthest but the closest one if GRBs did come from the hosts.
2, the fluences of the GRBs with the highest redshift all have a "**medium**" value instead of the least one.

Those are precisely contrary with the expectation of the cosmological origin theory of GRBs and the conclusions of quasars and hosts themselves. Moreover, Table 2 shows that, for the subset of GRB131, not only the values of both *b* and *r* are so small, but the signs are also incorrect. Therefore, we can confessedly affirm that there must be no affiliation between the distance of GRBs and the redshift of hosts.

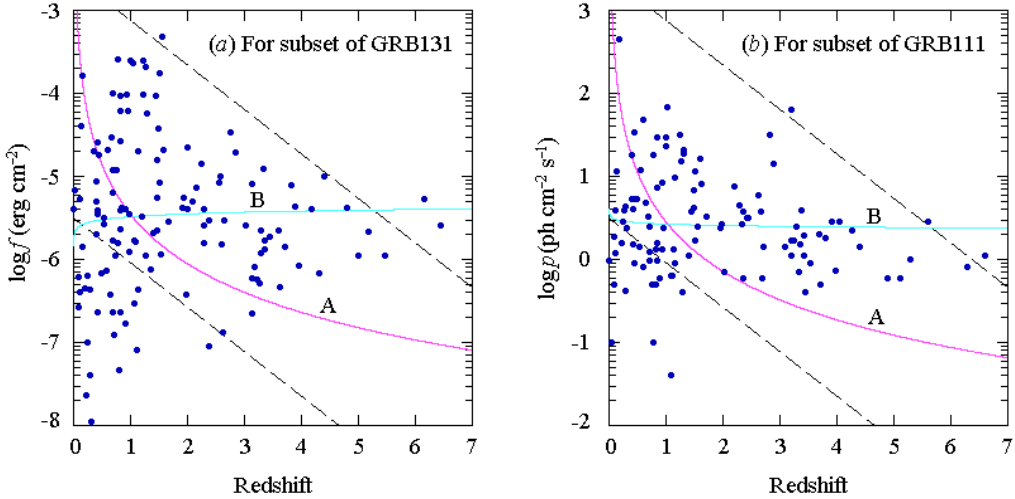

**FIG. 3.** Distributions of some apparent physical quantities of GRBs to the redshift; where curves A express that given by equations (4.1) or (4.2), curves B the regression curves. (*a*) The *f-z* relation for the subset of GRB131. (*b*) The *p-z* relation for the subset of GRB111.

FIG. 3(*b*) also shows that the faintest photon flux of GRBs are not the farthest but the closest one; there are total twenty-five samples (22.5%) in both the corners; on the other hand, the relevant values of *b* and *r* showing in Table 2 are so small that we can confessedly deny the affiliation between the distance of GRBs and the redshift of hosts again.

We have used two apparent physical quantities of GRBs to verify the correlation between the distance of GRBs and the redshift of hosts above, all of them gave the negative result. Therefore, we can affirm that there is no physical affiliation between the distance of GRBs and the redshift of hosts.

Since there is a strong correlation between the apparent magnitude of hosts and their redshift, well then whether there exists the possibility that the redshifts of hosts that *R*- or $R_c$- or *V*-magnitude was determined can also indicate the actual distance of GRBs there? In order to clarify the problem, we verify the correlation between the fluence and the redshift for the intersection of the subset of GRB131 and the subset of Host32 (called the subset of HF31 because which contains 31 GRBs). The distribution of the fluence *f* to the redshift *z* was shown in FIG. 4(*a*), and the results of correlation analysis were also shown in the Table 2. Comparing FIG. 2(*b*) and FIG. 4(*a*), likewise, comparing the relevant statistical results for the subsets of Host32 and HF31, we are sure that the situation and the conclusion have no any change; the redshift of hosts likewise can not indicate the distance of GRBs. Note, in special, that the subset of HF31 is a subset of Host32, the samples are almost selfsame in both FIG. 2(*b*) and FIG. 4(*a*), the difference between them is only the physical quantity: the fluence is only inherent in GRBs instead of hosts, and the apparent magnitude and redshift are only inherent in hosts instead of GRBs; the huge difference between two conclusions would categorically deny the physical affiliation between GRBs and hosts. The conclusion can be further confirmed by the fluence-apparent magnitude relation of the intersection of both subsets of GRB131 and Host32, i.e. the subset of HF31, showed in FIG. 4(*b*); which gives the correlation coefficient to be −0.0896, which sign is incorrect, and then shows that there is no correlation between the fluence of GRBs and the apparent magnitude of hosts, which is precisely opposite from FIG. 1.

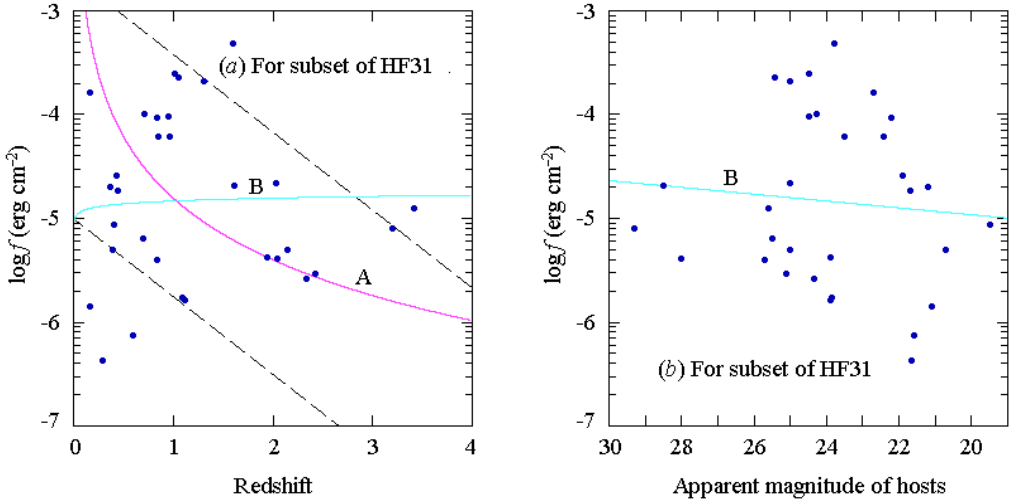

**FIG. 4.** Further verification of the correlation by subset of HF31. (*a*) The *f-z* relation for the subset of HF31, where curve A expresses the curve given by equation (4.1), curves B the regression curve. (*b*) Fluence-apparent magnitude relation for the subsets of HF31, B is the regression curve.

Moreover, we can further investigate the past records for the correlation between the redshift and the fluence (or the photon flux) of GRBs, or the apparent magnitude of hosts and the redshift. Let's put samples in order of the detected date, and do correlation analysis for each new sample. The correlation coefficients of each correlation analysis have been shown in FIG. 5.

FIG. 5(*a*) reveals the past records of the correlation coefficient between the fluence of GRBs and the redshift for the subset of GRB131; FIG. 5(*b*) reveals that between the photon flux of GRBs and the redshift for the subset of GRB111; FIG. 5(*c*) reveals that between the fluence of GRBs and the redshift for the subset of HF31; FIG. 5(*d*) reveals that between the apparent magnitude and the redshift for the subset of Host32. The historical records show clearly that there is no correlation between the

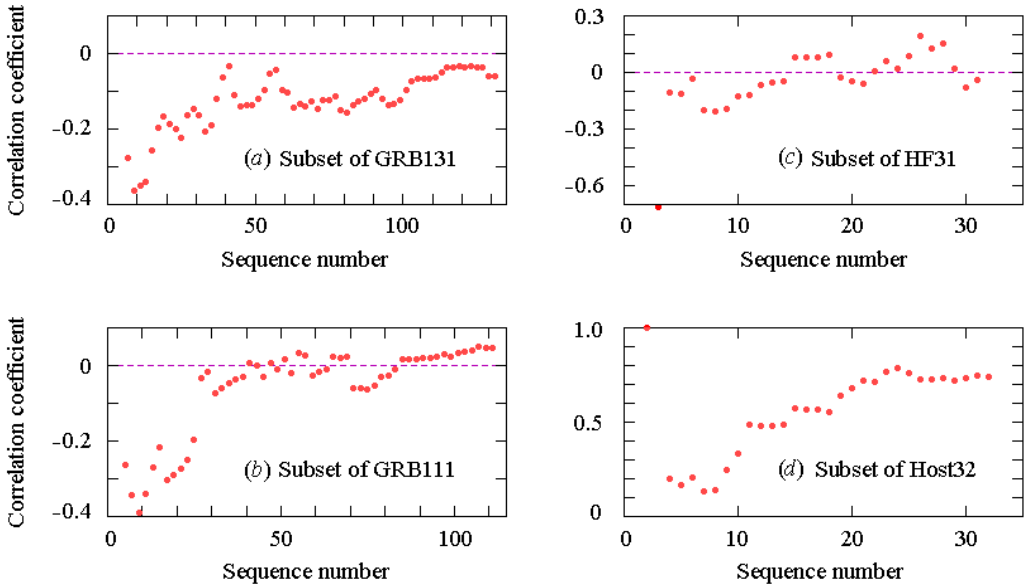

**FIG. 5.** Past records of correlation coefficient for different subsets. (*a*) Past records of correlation coefficient in *f-z* relation for the subset of GRB131. (*b*) Past records in *p-z* relation for the subset of GRB111. (*c*) Past records in *f-z* relation for the subset of HF31. (*d*) Past records in apparent magnitude-redshift relation for the subset of Host32.

apparent physical quantities of GRBs and the redshift of hosts although there is a stable significant correlation between the apparent magnitude of hosts and its redshift beginning from the tenth samples.

Therefore, we can lastly affirm that the distance of GRBs is independent from the redshift of hosts. In fact, the conclusion can be proved by a logic method, see below.

## 5. DEDUCTIVE REASONING FOR THE STATUS OF HOSTS

It is well known that the value of an apparent physical quantity would depend on the observers. For example, the apparent cross section of a nearby galaxy would vary inversely as its distance square and therefore depends on different observers located in different galaxies. However, an intrinsic physical quantity of a galaxy should be independent from the observer. For example, the probability of a galaxy generating a quasar is entirely independent from the observers.

Let's research the number-redshift relation for the closest quasars, which was showed in FIG. 6(*a*). It is well known that the number $N_{\text{gal}}(z)$ of nearby galaxies with redshift not exceeding $z$ is proportional to $z$ cube, i.e. $N_{\text{gal}}(z) \sim c_1 z^3$; we can therefore expect, if quasars indeed come from external galaxies, the number $N_{\text{quasar}}(z)$ of quasars with low redshift not exceeding $z$ should be also proportional to $z^3$. Indeed, from FIG. 6(*a*) we obtain that $N_{\text{quasar}}(z) = 62600\, z^3$ as expected for the closest 6000 quasars in the catalogue of dr5qso.dat. And then we obtain the probability of a nearby galaxy generating a quasar to be a constant and is independent from the observers. That is a reasonable conclusion because quasars are the genuine cosmological objects.

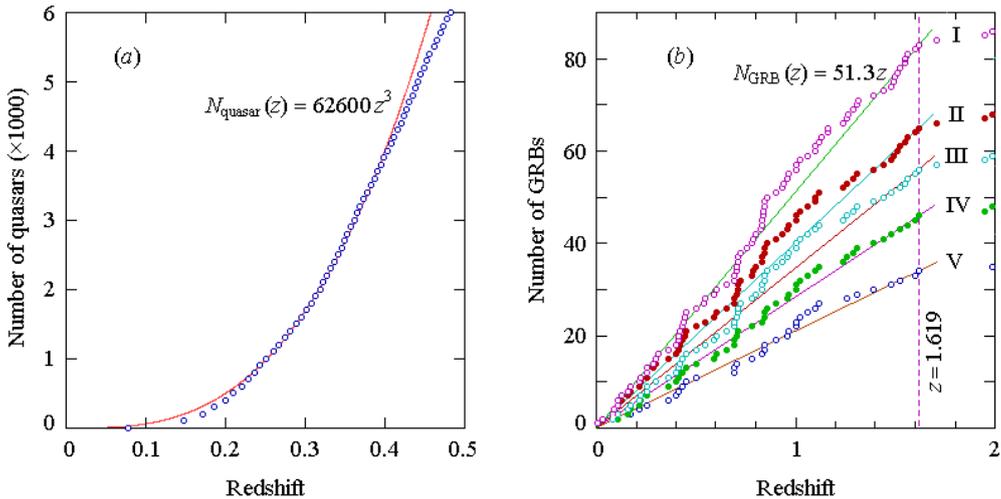

**FIG. 6.** Number-redshift relation for quasars and GRBs. (*a*) Number- redshift relation for the closest 6000 quasars. (*b*) Number-redshift relation for GRBs with redshift less than 2 and within the earliest 132, 105, 85, 65, 45 GRBs (from subsets I to V) in Table 1, respectively.

The redshift of the first 132 GRBs was determined in Table 1. FIG. 6(*b*) showed the number-redshift relations for those GRBs with redshift not exceeding 2 within different subsets among the first 132 GRBs. The graph showed an evident observed law that *the number $N_{GRB}(z)$ of GRBs with redshift not exceeding z is proportional to z in the low redshift region*:

$$N_{GRB}(z) \cong Cz. \quad (5.1)$$

The linear relation showed in equation (5.1) is stable, which is independent from the number of GRBs. For example, the linear relation will remain valid no matter for entire 132 GRBs, or for the earliest 105 GRBs (from GRB970228 to GRB060707), the earliest 85 GRBs (from GRB970228 to GRB060116), the earliest 65 GRBs (from GRB970228 to GRB050730), or earliest 45 hosts (from GRB970228 to GRB040827) in Table 1, and there are $C \cong 51.3$, 40.1, 34.5, 28.4 and 21.0 respectively.

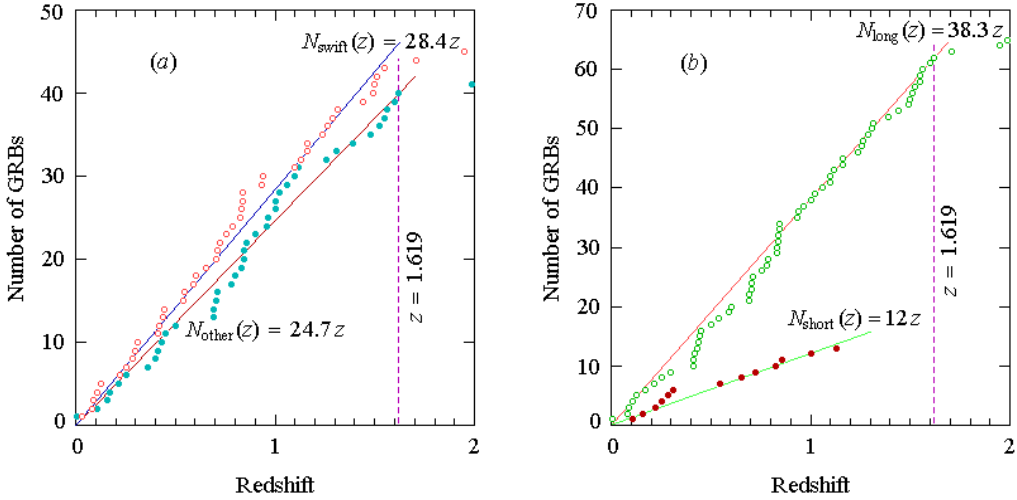

**FIG. 7.** Number-redshift relations for GRBs. (*a*) Number-redshift relation for GRBs detected by the Swift and other satellites, respectively. (*b*) Number-redshift relation for "long" bursts and "short" bursts, respectively.

The linear relation will likewise remain valid for other random subsets of GRBs, e.g. no matter for the samples detected by either the Swift or other satellites [see also FIG. 7(*a*)], or for both the "long" bursts and the "short" bursts [see also FIG. 7(*b*)]; even though for both the subsets of GRB131 and GRB111, the linear relation also remains valid.

Using the observed law of the number-redshift relation of GRBs, we can prove that GRBs do not come from the host by the reduction to absurdity.

Due to $N_{\text{gal}}(z) \sim c_1 z^3$ for the nearby galaxies, from equation (5.1) we can deduce a probability function in the low redshift region:

$$P_{\text{bd}}(z) = \frac{dN_{\text{GRB}}(z)}{dN_{\text{gal}}(z)} = \frac{c_0}{z^2}. \qquad (5.2)$$

Notice that the distance of a nearby galaxy is proportional to its redshift, we then obtain that *the probability of a nearby galaxy becoming a host of a GRB would vary inversely as its distance square if GRBs indeed come from the external galaxies.*

From the law, we can easily prove that GRBs do not come from the host. Indeed, if GRBs were generated in the host galaxy, then the probability of a galaxy *becoming* a host should be selfsame with that of a galaxy *generating* a GRB, and then the conclusion that "the probability of a nearby galaxy *generating* a GRB varies inversely as its distance square" would be bound to obtain. However, that is clearly a wrong one because the probability has become an apparent physical quantity instead of the intrinsic physical quantity now. Let's consider two galaxies I and II lying in a sight line and with distance of $2 \times 10^{21}$ miles (corresponding to the redshift about 0.02) and $6 \times 10^{21}$ miles respectively. We obtain, from equation (5.2), that the probability of galaxy I is nine times as much as that of galaxy II for us. Furthermore, let's use the formula (5.2) to the observers located in different galaxies. Relative to some observers located in some other galaxies, both galaxies I and II would have an identical distance, according to equation (5.2), their probabilities must be selfsame; while relative to another observer located in another galaxy, the distances of galaxy II and I would be $10^{21}$ miles and $5 \times 10^{21}$ miles, respectively, according to equation (5.2), the probability of galaxy II must be twenty-five times as much as that of galaxy I; and so on. In this way, the theory has already lost the self-consistency here: the probability of a galaxy *generating* a GRB is essentially an intrinsic physical quantity that is entirely independent from the observers; nevertheless, it has become an apparent physical quantity depending on the observers now and that is undoubtedly wrong. Notice that equation (5.2) is an inference deduced from an observed law, there is no problem in itself. The problem comes from the conclusion that the redshift is inherent in the host, and which has been proved to be wrong above. Therefore, GRBs do impossibly come from the host; the remainder possibility is that the redshift either originates from the background galaxy or results from the gravity; the third possibility does not exist.

The conclusion regarding the status of hosts would be valid under a broad observed condition. In fact, as long as the number-redshift observed law of GRBs satisfies $N_{\text{host}} = c_2 z^\alpha$ and $\alpha \neq 3$, it can be likewise proved that GRBs do not come from the host galaxy.

Can the theory explain the observed law of GRBs in the case of background galaxy? The answer is affirmative because the probability of a galaxy becoming a background galaxy is proportional to its apparent cross section and which varies inversely as its distance square, i.e. the probability varies inversely as its distance square.

In fact, the precise background galaxy theory can interpret the observed laws of GRBs, see §7 below.

The linear relation of the number-redshift distribution of GRBs has extended to $z \geq 1$ even higher $z$, see also FIG. 6($b$), FIG. 7($a$) and FIG. 7($b$) above; this is a convincingly collateral evidence for demonstrating that GRBs do not come from the host galaxy. It is well known that the influence from the evolutionary effects and the spatial curvature would be expected for the distribution of faraway objects; FIG. 6($a$) shows that the influences had in fact appeared evidently from $z \geq 0.4$ because the number-redshift relation of quasars has deviated from $N_{quasar}(z) = 62600\, z^3$. If GRBs indeed come from the depth of the Universe as indicated by the redshift, we can then expect that the number-redshift relation of GRBs would show some evident change when $z > 0.4$; while the expectation has no occurred, the linear relation continuously extended to $z \geq 1$ even higher $z$; which cannot be interpreted by the cosmic GRB theory.

In order to find other collateral evidences, we consider the threshold of the instruments. It is well known that there are detection thresholds of the detectors in the detection of GRBs, every detector has its lowest detection limit; a GRB would not be detected if its peak photon flux is lower than the thresholds. This effect is called as "flux-limited (effect)" here.

Does the flux-limited indeed affect the number-redshift relation for GRBs? The answer is negative. FIG. 8($a$) showed the results for following four kinds of flux-limited effects:

1. For all GRBs with $p \geq 16$ ph cm$^{-2}$ s$^{-1}$, their detection efficiency is 100%, i.e. there is no flux-limited effect for them because the detection efficiency is all 100% when 1 s peak photon flux of a GRB satisfies $p \geq 1.562$ ph cm$^{-2}$ s$^{-1}$ (based on BATSE Criteria).

2. For all GRBs with $p \geq 4.3$ ph cm$^{-2}$ s$^{-1}$, their detection efficiency is also 100%, i.e. there is no flux-limited effect for them.

3. For all GRBs with $p \geq 1.6$ ph cm$^{-2}$ s$^{-1}$, their detection efficiency is also 100%, i.e. there is no flux-limited effect for them.

4. For all 111 GRBs which 1 s peak photon fluxes are all determined and there would exist the flux-limited effect if it indeed exists.

FIG. 8($b$) showed the number-redshift relation corresponding to the four kinds of flux-limited effects above.

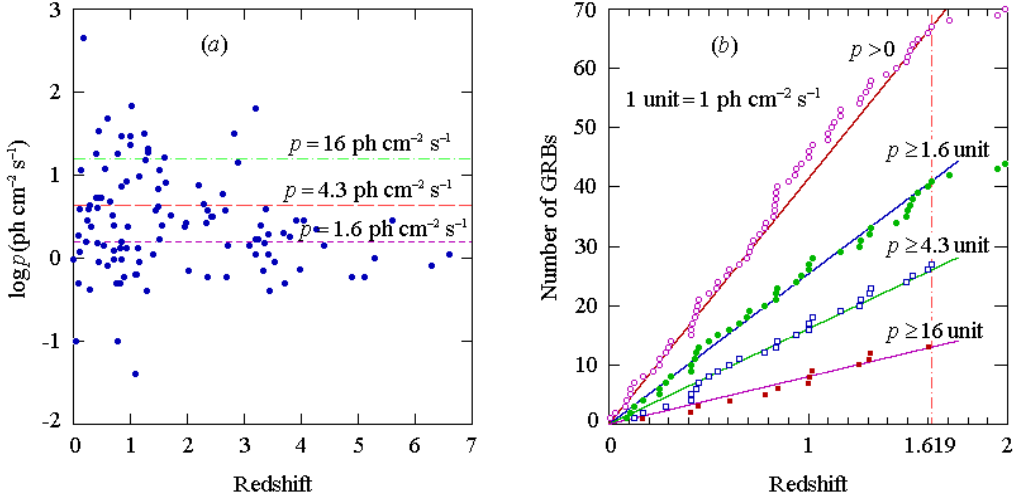

**FIG. 8.** Non-flux-limited effect for GRBs. (*a*) Selection for the flux limit of GRBs with different cut-off fluxes. (*b*) Number-redshift relation for both GRBs that has flux-limited effect and have no flux-limited effect.

From FIG. 8(*b*) we obtain the following conclusions:

1. The number-redshift relation of GRBs does not be affected by the so called "flux-limited (effect)" because the linear relationship for four subsets of GRBs showing in FIG. 8(*b*) is in effect selfsame.

2. The linear distribution relation of a random subset of GRBs is independent from the sample size.

## 6. PREDICAMENTS FROM THE LUMINOSITIES

It would be shown below that as long as the redshift is thought to indicate the distance of GRBs, more inconsistent conclusions would be deduced. As mentioned above, there is no correlation between the redshift and both the fluence and the peak photon flux of GRBs. If we compel the redshift to indicate the distance of GRBs, then, according to equation (2.1), the luminosities $F$ and $P$ must be averagely proportional to the distance square because it has been proved above that the values of both $f$ and $p$ are independent from the redshift. Indeed, according to equations (2.1) and (2.2), we obtain the distribution of luminosity $P$ of the photon flux for the subset of GRB111 as showed in FIG. 9(*a*) and that of luminosity $F$ of the fluence for the subset of GRB131 as showed in FIG. 9(*b*). Here, for convenience, we put $d = 1$, $P = 1$ and $F = 1$ for the closest GRB (i.e. the GRB980425, $z = 0.0085$).

FIG. 9(*a*) indeed shows that the mean luminosity $\bar{P}$ of the peak photon flux

varies as the distance square no matter the distance is far or near; likewise, FIG. 9(*b*) shows that the mean luminosity $\overline{F}$ of the fluence also varies as the distance square no matter the distance is far or near. The green dot lines in FIG. 9(*a*) and FIG. 9(*b*) show the linear regression results, which give $\overline{P} = 3.666 d^{1.941}$ and $\overline{F} = 0.470 d^{2.116}$; where the other lines have slope of 2. Thereupon, the theory has returned to the inconsistent state again because the luminosities of GRBs have become a very strange apparent physical quantity that depends on the observers now: the farther the observer, the greater the values of the luminosities will be; the closer the observer, the less the values of the luminosities will be; which entirely violates the basic principle of physics. Note that these are not caused by the evolutionary effects or the spatial curvature, because the luminosities of the closer GRBs (e.g. $z \leq 0.4$) are also to show this behavior, see also FIG. 9(*a*) and FIG. 9(*b*). Therefore, the luminosities of GRBs have become an unusual apparent physical quantity now and that is incorrect.

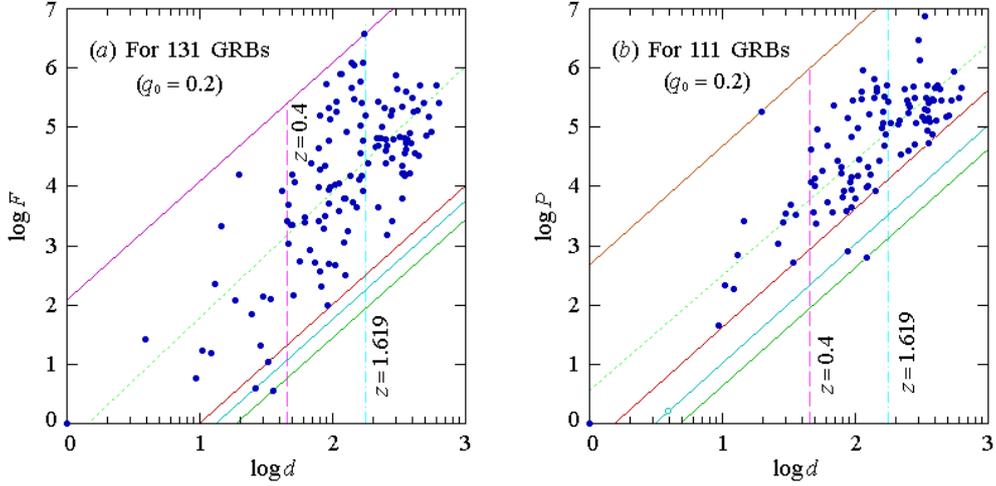

**FIG. 9.** Luminosity-distance relation for GRBs. (*a*) Luminosity-distance relation of the photon flux for the subset of GRB111. (*b*) Luminosity-distance relation of the fluence for the subset of GRB131.

On the other hand, we found, from FIG. 9(*a*) and 9(*b*), that the distribution amplitude of both luminosities *P* and *F* have reached seven orders of magnitude [log $(P_{max}/P_{min}) = 6.932$, log $(F_{max}/F_{min}) = 6.604$], and $d^2$ reached six orders of magnitude [log $(d_{max}/d_{min})^2 = 5.703$]; in general, we can expect that the distribution amplitudes of both *p* and *f* would reach ten orders of magnitude or higher! Nevertheless, the actual distribution amplitudes of both *p* and *f* so much as do not reach five orders of magnitude [log $(p_{max}/p_{min}) = 4.052$, log $(f_{max}/f_{min}) = 4.646$], which are much less than that of *P* and *F* and so much as less than that of $d^2$! This is a very unreasonable result and which entirely arises from the incorrect idea that GRBs distance depends on the redshift.

## 7. BACKGROUND GALAXY THEORY

Under the major premise that hosts of GRBs are entirely the background galaxy, all physical quantity distribution laws of GRBs to the redshift of background galaxies can be deduced theoretically, which can be accurately verified by the observed data. We found the theory in a standard Friedman's cosmology as below (Weinberg 1972).

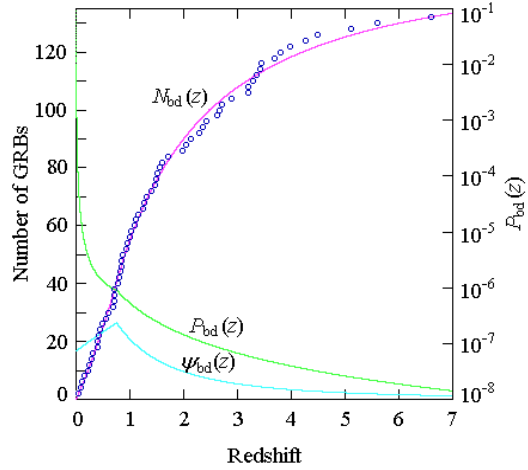

**FIG. 10.** Verification for the background galaxy theory.

According to the big bang theory, the Universe evolves from small to big, the size of the early universe (and galaxies) is small. The mean cross section $S = S(\tau)$ of galaxies should be a function of its cosmological age $\tau$, which would increase as its age growth, but it cannot increase infinitely. Hence, we can assume $S(\tau)$ to be

$$S(\tau) = \begin{cases} k\tau^\alpha, & \text{when } \tau < \tau_0; \\ k\tau_0^\alpha, & \text{when } \tau \geq \tau_0. \end{cases} \quad (7.1)$$

Where, $k$, $\alpha$ and $\tau_0$ are undetermined parameters. The field angle of the section $S(\tau)$ for current observation is $S(\tau)/d_A^2$, where $d_A$ is the angular diameter distance:

$$d_A = \frac{c\left(q_0 z + (q_0-1)\left\{\sqrt{1+2q_0 z}-1\right\}\right)}{H_0 q_0^2 (1+z)^2}. \quad (7.2)$$

The probability of a galaxy with cross section $S(\tau)$ becoming a background galaxy is equal to

$$P_{bd}(q_0, z) = \frac{S(\tau)}{4\pi d_A^2} = \begin{cases} \dfrac{c_0 q_0^4 (1+z)^4}{\left\{q_0 z + (q_0-1)\left(\sqrt{1+2q_0 z}-1\right)\right\}^2}, & \text{when } z \leq z_0; \\ \dfrac{c_0 q_0^4 (1+z)^4}{\left\{q_0 z + (q_0-1)\left(\sqrt{1+2q_0 z}-1\right)\right\}^2}\left(\dfrac{\tau}{\tau_0}\right)^\alpha, & \text{when } z > z_0. \end{cases} \quad (7.3)$$

Where, $c_0$ is the normalization constant. The curve $P_{bd}(z) = P_{bd}(q_0, z)$ was shown in FIG. 10 (in a relative unit with $\alpha = 2$ and $q_0 = 0.148$, see below). For lower $z$ we have

$$P_{bd}(q_0, z) \cong \frac{c_0}{z^2},$$

which is precisely the equation (5.2) —the second statistical law of GRBs.

Let $dN(q_0, z)$ be the number of galaxies that redshift is located in $z$ to $z + dz$,

$$dN(q_0,z) = C_1 \frac{\left(q_0 z + (q_0-1)\left\{\sqrt{1+2q_0 z}-1\right\}\right)^2}{q_0^4 (1+z)^3 \sqrt{1+2q_0 z}} dz. \tag{7.4}$$

Where, $C_1$ is a constant. The total field angle of all $dN(q_0, z)$ galaxies is equal to

$$d\Omega = \frac{S(\tau)}{d_A^2} dN(q_0,z) = \frac{H_0^2 C_1 S(\tau)}{c^2} \frac{(1+z)}{\sqrt{1+2q_0 z}} dz.$$

Hence, the total probability for $dN(q_0, z)$ galaxies is equal to

$$dp(q_0,z) = \frac{d\Omega}{4\pi} = \begin{cases} C \dfrac{1+z}{\sqrt{1+2q_0 z}} dz, & \text{when } z \leq z_0; \\ C \dfrac{1+z}{\sqrt{1+2q_0 z}} \left(\dfrac{\tau}{\tau_0}\right)^\alpha dz, & \text{when } z > z_0. \end{cases} \tag{7.5}$$

Where, $C = kH_0^2 C_1 \tau_0^\alpha / 4\pi c^2$ is the normalization constant.

For total $N_0$ background galaxies, the expectation of the number of background galaxies with redshift not exceeding $z$ is as follows:

$$N_{\text{bd}}(q_0, z) = \begin{cases} CN_0 \displaystyle\int_0^z \dfrac{(1+z)}{\sqrt{1+2q_0 z}} dz, & z \leq z_0; \\ N_{\text{bd}}(q_0, z_0) + CN_0 \displaystyle\int_{z_0}^z \dfrac{(1+z)(\tau/\tau_0)^\alpha}{\sqrt{1+2q_0 z}} dz, & z > z_0. \end{cases} \tag{7.6}$$

This is the number-redshift relation for the background galaxies of GRBs and the curve $N_{\text{bd}}(z) = N_{\text{bd}}(q_0, z)$ was shown in FIG. 10. Where, let

$$x_0 = \frac{1-2q_0}{2q_0(1+z_0)}, \quad x = \frac{1-2q_0}{2q_0(1+z)}$$

when $q_0 < 0.5$, we have

$$\frac{\tau}{\tau_0} = \frac{\sqrt{x(1+x)} - \log\left(\sqrt{x} + \sqrt{1+x}\right)}{\sqrt{x_0(1+x_0)} - \log\left(\sqrt{x_0} + \sqrt{1+x_0}\right)}.$$

Equation (7.6) can be integrated when $z \leq z_0$:

$$N_{\text{bd}}(q_0,z) = C \frac{2z}{1+\sqrt{1+2q_0 z}} \left(1 + \frac{z}{3}\left(1 + \frac{z}{1+\sqrt{1+2q_0 z}}\right)\right)$$

$$\cong C\left(z + \frac{(1-q_0)}{2} z^2 + O(z^3)\right). \tag{7.7}$$

We then obtain that the number of nearby background galaxies is proportional to the redshift, which is precisely the first statistical law of GRBs showing in equation (5.1).

From equation (7.5), we can obtain a differential probability function as follows:

$$\psi(q_0, z) = \frac{dp(q_0, z)}{dz} = \begin{cases} \dfrac{C(1+z)}{\sqrt{1+2q_0 z}}, & \text{when } z \leq z_0; \\ \dfrac{C(1+z)}{\sqrt{1+2q_0 z}} \left(\dfrac{\tau}{\tau_0}\right)^{\alpha}, & \text{when } z > z_0. \end{cases} \quad (7.8)$$

The curve $\psi_{bd}(z) = \psi(q_0, z)$ has been shown in FIG. 10 (in a relative unit).

The theory has provided a special method for determining the deceleration parameter $q_0$ of the Universe here; that is a best method with the least uncertain factors because there are only three undetermined parameters $q_0$, $\alpha$ and $z_0$ in $N_{bd}(q_0, z)$ with a constraint that $q_0$, $\alpha$ and $z_0$ must ensure $N_{bd}(q_0, z)$ to coincide with the practical number-redshift relation of background galaxies well.

In order to obey the constraint, parameters $q_0$, $\alpha$ and $z_0$ must be carefully adjusted. We find the best result with $z_0 = 0.755$, $\alpha = 2$, $q_0 = 0.148$ and $C = 0.3125$. This is a reasonable one because the fact of $\alpha = 2$ implies that the diameter of a galaxy is proportional to its cosmological age in the early cosmos. The result was shown as the curve $N_{bd}(z) = N_{bd}(q_0, z)$ in FIG. 10. As a preliminary estimation, we then obtain $q_0 = 0.148$ and the Universe density of $\rho_{matter} = 0.296 \rho_c$ which has evidently contained the contribution of the dark matter because $q_0$ within $N_{bd}(q_0, z)$ is the parameter that reflect the spatial curvature which must contain the contribution of the dark matter.

And then the theory has successfully explained all the observed laws of GRBs. It is obvious that the probability of a higher redshift galaxy becoming a background galaxy is very small.

The method that determines the deceleration parameter by using the redshift distribution of the background galaxy is not limited to the observation of GRBs only, but also used for other observations. For example, we can use the redshift distribution of the background galaxy behind the point of intersection of two diagonals of a quadrilateral consisting of four Galactic stars to determine the deceleration parameter, which would be much better than the observation of GRBs.

## 8. VARIABLES SEPARATION TEST

For confirming whether GRBs do come from the host or not, except for the correlation statistical analysis and the deductive reasoning, the third method is the variables separation test. If the apparent physical quantity and the redshift of GRBs can be separated from the distribution function, we can affirm that GRBs do not come from the host. Let us introduce the distribution function $\psi(\log f, z)$ for the fluence $f$ [or $\psi(\log p, z)$ for the photon flux $p$]; if RGBs indeed come from the host, then the distance $d$ of GRBs must depend on the redshift $z$: $d = d(z)$, and then the $\log f$ (or $\log p$) of course cannot be separated from the redshift $z$ in the distribution function because there must be $\log f = \log(F/4\pi) - 2\log d(z)$ now; on the contrary, if $d$ is independent from $z$, then $\log f$ (or $\log p$) can definitely be separated from $z$ in the distribution function. We then obtain $\psi(\log f, z) = \varphi(\log f)\psi(z)$ in the later case, where, $\psi(z)$ is the probability distribution function of the redshift variable $z$ of GRBs and $\varphi(\log f)$ is the distribution function of the fluence variable $f$ of GRBs; and so on.

How identify whether the quantities $f$ and $z$ can be separated from $\psi(\log f, z)$? This is in fact very simple. It is well known that each physical quantity $w$ of GRBs would have a distribution function $\varphi(w)$ and $\varphi(w)$ would generally have a maximum that corresponds to the most probable value of $w$; the physical quantity $w$ of the most GRBs would densely distribute around the most probable value as mentioned above. Notice that $\psi(\log f, z)$ is the probability of the GRB event with physical quantity $f$ and redshift $z$; when $\log f$ and $z$ can be separated, we have $\psi(\log f, z) = \varphi(\log f)\psi(z)$, where $\psi(z)$ is the probability that GRB redshift takes $z$. Since the sample size of GRBs is in general smaller, we can then expect that only the event with not too small probability $\psi(\log f, z)$ can occur. As mentioned above, the probability occurring a GRB with higher redshift is very small, therefore, if a GRB occurs with a higher redshift, the value of $\psi(z)$ must be very small; in order to ensure such event can occur, $\varphi(\log f)$ must close to the maximum such that $\psi(\log f, z)$ would be not too small as mentioned in Rule 3 of §2 above. Therefore, there must be an evident distribution characteristic if GRBs distance is independent from the redshift: *the values of the physical quantity* $\log f$ *and* $\log p$ *of GRBs should close to the most probable value such that* $\varphi(\log f)$ *and* $\varphi(\log p)$ *closes to the maximum when GRB distance is independent from the redshift*. This is called "*the independent physical quantity law*".

However, the independent physical quantity law is not valid for the objects that distance depends on the redshift, e.g. the quasars and the hosts, in which the apparent magnitude $w$ and $z$ within function $\psi(w, z)$ cannot be separated because there must be $w = w_0 + 5\log \phi(q_0, z)$ there; the distribution function $\psi(w, z)$ therefore cannot be written as $\psi(w, z) = \varphi(w)\psi(z)$ although we can define functions $\varphi(w)$ and $\psi(z)$ as follows:

$$\varphi(w) = \int_0^\infty \psi(w, z)dz, \quad \psi(z) = \int_0^\infty \psi(w, z)dw,$$

nevertheless, there must be $\psi(w, z) \neq \varphi(w)\psi(z)$ here. This is precisely the third method for determining whether there is a physical affiliation between the GRB distance and the redsgift.

In order to verify the law, the distribution function $\varphi(\log f)$ or $\varphi(\log p)$ of GRBs must be determined. The empiric distribution function for each subset of GRBs can easily be obtained by graphic method.

FIG. 11($a$) showed the distribution function $\varphi(\log f)$ of the fluence for the subset of GRB131; from FIG. 11($a$) we obtain the distribution function $\varphi(\log f)$ as follows:

$$\varphi(x) = 0.025\exp[-1.2(x+8)^2] + 0.1\exp[-2(x+3.9)^2]$$
$$+ 0.475\exp[-0.95(x+5.62)^2]. \qquad (8.1)$$

Where, $x = \log f$; the maximum of $\varphi(\log f)$ is at $\log f = -5.6182$.

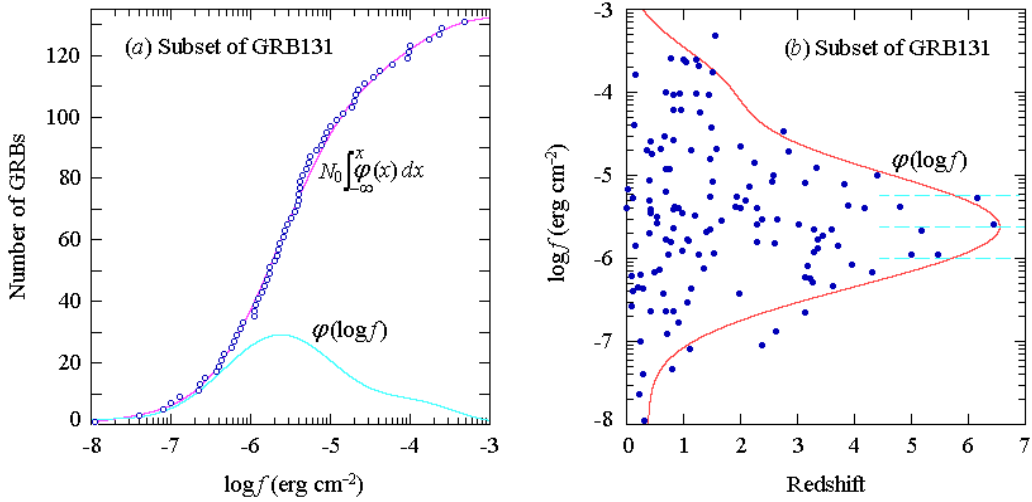

**FIG. 11.** Distribution function and verification. ($a$) Distribution function $\varphi(\log f)$ of the fluence for the subset of GRB131. ($b$) Verification for the independent physical quantity law.

FIG. 11($b$) showed the verification for the independent physical quantity law. From FIG. 11($b$) we found that the six GRBs with redshift $z > 4.5$ indeed all appear at the region closed to the maximum of the distribution function $\varphi(\log f)$ as expected by the independent physical quantity law; this means that the physical quantity $\log f$ can be separated from $z$ in the distribution function $\psi(\log f, z)$ and then the distance of GRBs is really independent from the redshift.

FIG. 12($a$) showed the distribution function $\varphi(\log p)$ of the 1 s peak photon flux for the subset of GRB111; from FIG. 12($a$) we obtain the distribution function $\varphi(\log p)$ as follows:

$$\varphi(x) = 0.035\exp[-8(x+1.4)^2] + 0.167\exp[-8(x-1.43)^2]$$
$$+ c(x+0.55)\exp[-(x+0.55)^2]. \qquad (8.2)$$

Where, $x = \log p$; $c = 0$ if $x < -0.55$, otherwise $c_2 = 1.75$; the maximum of $\varphi(\log p)$ is at $\log p = 0.157$.

FIG. 12($b$) showed the verification for the independent physical quantity law. From FIG. 12($b$) we found that the twelve GRBs with the highest redshift indeed all very close to the maximum of the distribution function $\varphi(\log p)$ as expected by the

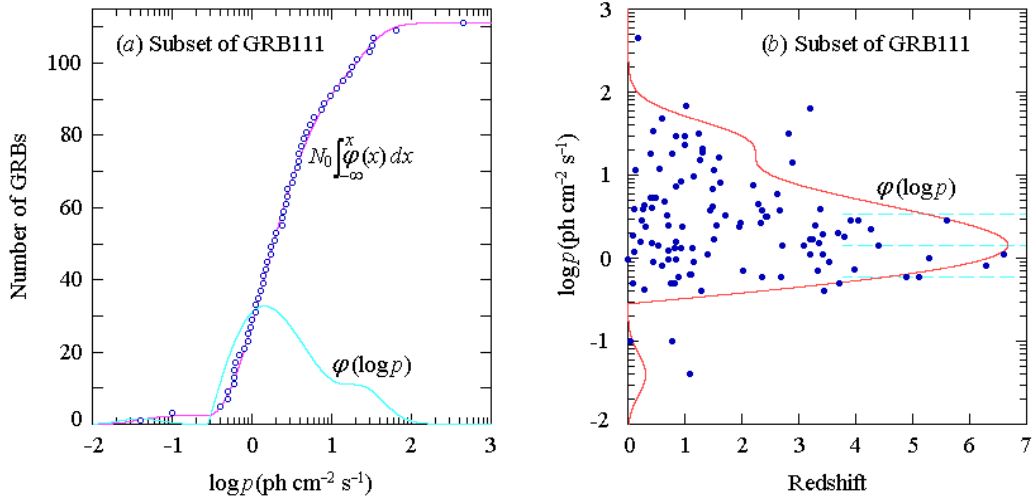

**FIG. 12.** Distribution function and verification. (*a*) Distribution function $\varphi(\log p)$ of the 1 s peak photon flux for the subset of GRB111. (*b*) Verification for the independent physical quantity law.

independent physical quantity law; this also implies that physical quantity $\log p$ can be separated from $z$ in the distribution function $\psi(\log p, z)$ and then the distance of GRBs is really independent from the redshift.

FIG. 13(*a*) showed the distribution function $\varphi(\log f)$ of the fluence for the subset of HF31; from FIG. 13(*a*) we obtain the distribution function $\varphi(\log f)$ as follows:

$$\varphi(x) = 0.485\exp[-1.1(x+5.16)^2] + 0.31\exp[-8(x+3.75)^2]. \tag{8.3}$$

Where, $x = \log f$.

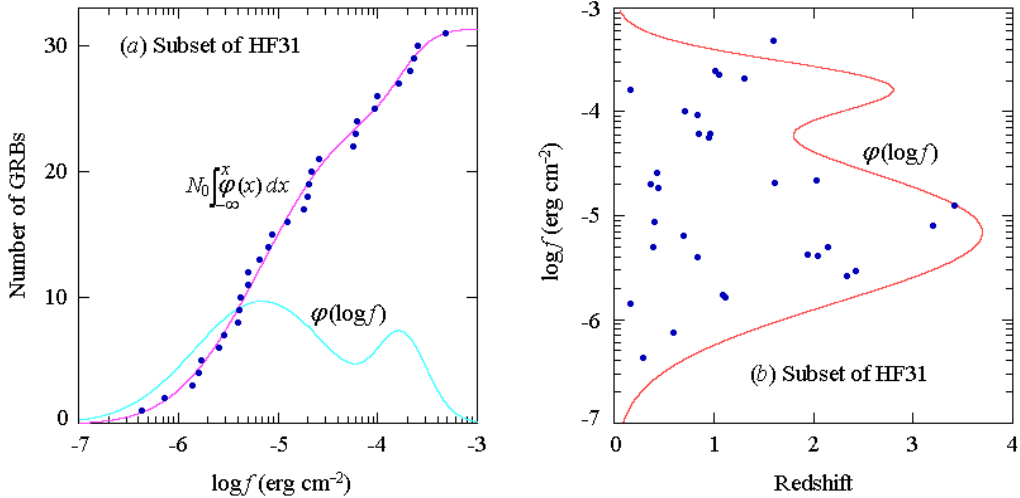

**FIG. 13.** Distribution function and verification. (*a*) Distribution function $\varphi(\log f)$ of the fluence for the subset of HF31; (*b*) Verification for the independent physical quantity law.

FIG. 13(*b*) showed the verification of the independent physical quantity law. From FIG. 13(*b*) we found that the eight GRBs with the highest redshift do indeed appear at the region closed to the maximum of the distribution function $\varphi(\log f)$ as expected by the independent physical quantity law; this also means that the physical quantity $\log f$ can be separated from $z$ in the distribution function $\psi(\log f, z)$ and then the distance of GRBs is really independent from the redshift.

Now, we demonstrate that the independent physical quantity law is not valid for the objects that apparent physical quantity depends on the redshift such as both the quasars and the hosts.

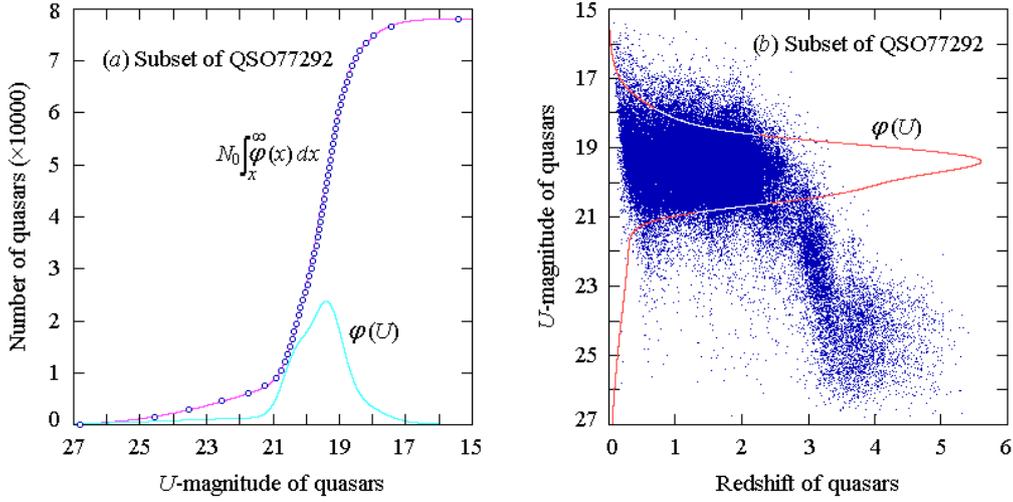

**FIG. 14.** Distribution function and verification. (*a*) Distribution function $\varphi(U)$ of the U-magnitude for the subset of QSO77292; (*b*) Verification for the independent physical quantity law.

FIG. 14(*a*) showed the distribution function $\varphi(U)$ of the *U*-magnitude for the subset of QSO77292 obtained by

$$\varphi(U) = \int_0^\infty \psi(U, z) dz,$$

From FIG. 14(*a*) we obtain the distribution function $\varphi(U)$ as follows:

$$\varphi(U) = 0.00227(15-U)^2 \exp[-0.03(15-U)^2] + 0.2323 \exp[-3(20.3-U)^2]$$
$$+ 0.4123 \exp[-2.5(19.37-U)^2] + 0.07 \exp[-(18.55-U)^2]. \qquad (8.4)$$

FIG. 14(*b*) showed the verification of the independent physical quantity law for quasars. From FIG. 14(*b*) we found that all the quasars with the highest redshift do not appear at the region closed to the maximum of the distribution function $\varphi(U)$ but far away from it. This means that $\varphi(U)$ can only indicate the dense position of quasars but not indicate the position of the samples with highest redshift. This is inevitable

because the physical quantity $U$ depends on the redshift $z$ and they cannot be separated from the distribution function $\psi(U, z)$; therefore, the independent physical quantity law is invalid for the quasars. This shows that the method for determining whether there is a physical affiliation between the distance and the redshift is valid.

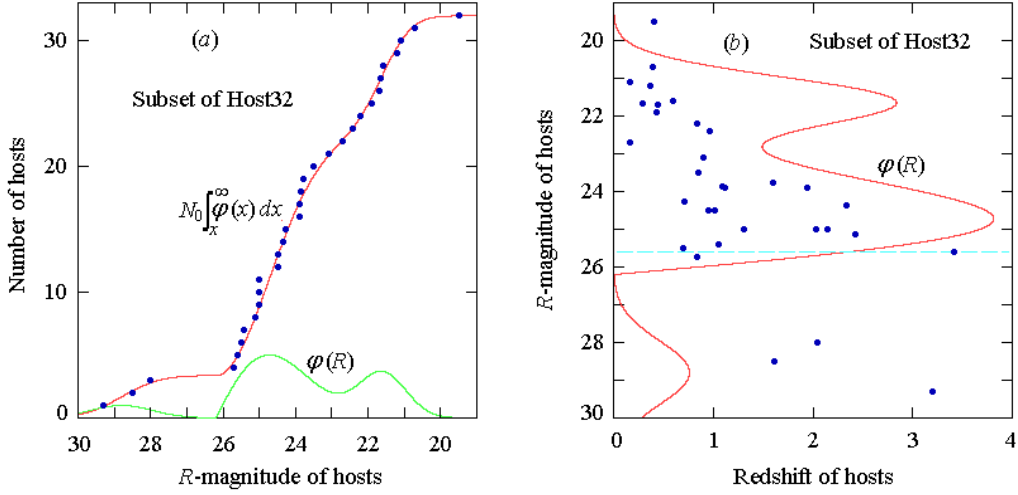

**FIG. 15.** Distribution function and verification. (*a*) Distribution function $\varphi(R)$ of the apparent magnitude for the subset of Host32. (*b*) Verification for the independent physical quantity law.

FIG. 15(*a*) showed the distribution function $\varphi(R)$ of the apparent magnitude for the subset of Host32 obtained by

$$\varphi(R) = \int_0^\infty \psi(R,z) dz ,$$

From FIG. 15(*a*) we obtain the distribution function $\varphi(R)$ as follows:

$$\varphi(R) = 0.05\exp[-0.7(28.8 - R)^2] + 0.176\exp[-1.2(21.6 - R)^2]$$
$$+ c(26.2 - R)\exp[-0.23(26.2 - R)^2] . \qquad (8.5)$$

Where, $c = 0$ if $R > 26.2$, otherwise $c = 0.28$.

FIG. 15(*b*) showed the verification of the independent physical quantity law for hosts. From FIG. 15(*b*) we found that there are two peaks with almost the same height in the distribution function $\varphi(R)$; there is no sample with highest redshift appearing at the nearby region of the second peak; although there is a sample with the highest redshift at the nearby region of the first peak, but the value of $\varphi(R)$ is only *medium*; the another sample with the highest redshift does straightforwardly appear at the low value region of $\varphi(R)$. This means that $\varphi(R)$ can only indicate the dense position of hosts but does not indicate the position of the samples with highest redshift; in other words, the independent physical quantity law is invalid for hosts. This is inevitable because the physical quantity $R$ depends on the redshift $z$ and they cannot be separated

from the distribution function $\psi(R, z)$.

In a word, we can affirm that the distance of GRBs is independent from the redshift in terms of the independent physical quantity law.

## 9. PREDICTION AND VERIFICATIONS

The distribution of the apparent physical quantity of GRBs is generally dispersed in a broad region up to five orders of magnitude. It is in general impossible that forecast the value of an apparent physical quantity for a GRB that will soon be detected. However, there is an exception for those GRBs with the highest redshift.

According to the independent physical quantity law, we can forecast the values of certain physical quantities for those GRBs with higher redshift; here, the redshift must close or exceed a definite value.

In order to show the reliability of this method, let's seek the foundation from the past records. The first GRB with redshift exceeding 6.0 is GRB050904; in order to forecast the values of the fluence and the photon flux for those GRBs that redshift exceeds or closes to 6.0, let's draw the *f-z* distribution figure and the *p-z* distribution figure for those GRBs detected earlier than GRB050904.

FIG. 16(*a*) showed the distribution function $\varphi(\log f)$ of the fluence for the earliest 71 GRBs that detected earlier than GRB050904; from FIG. 16(*a*) we obtain

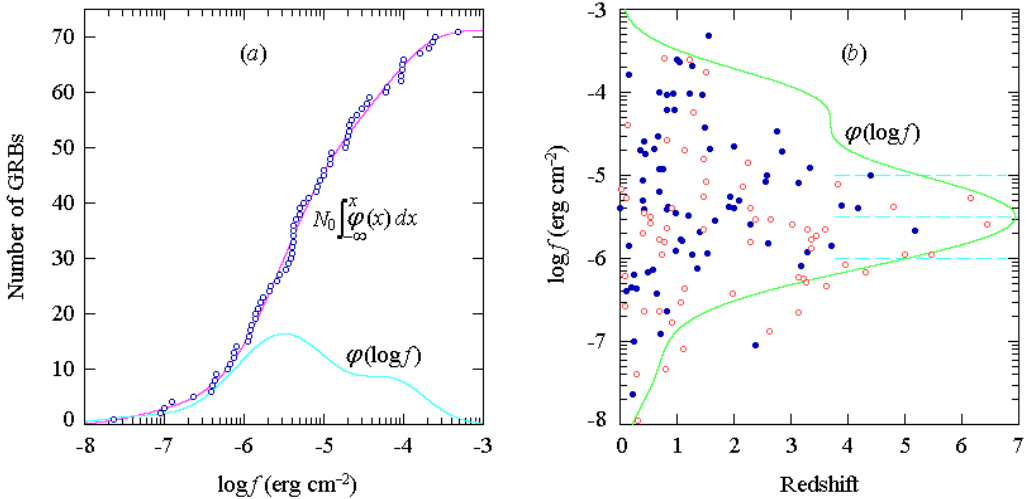

**FIG. 16.** Distribution function and forecast. (*a*) Distribution function $\varphi(\log f)$ for 71 GRBs detected earlier than GRB050904. (*b*) The *f-z* relation of 131 GRBs; where, the solid circles express the 71 GRBs detected earlier than GRB050904; the open circles express the others.

the distribution function of $\varphi(\log f)$ as follows:

$$\varphi(x) = 0.040\exp[-2(x+7.3)^2] + 0.480\exp[-1.3(x+5.5)^2]$$
$$+ 0.205\exp[-3(x+4.1)^2]. \quad (9.1)$$

Where, $x = \log f$; the maximum of $\varphi(\log f)$ is at $\log f = -5.4964$.

FIG. 16(*b*) showed the $\log f$–$z$ relation for whole 131 GRBs, in which the solid circles express the 71 GRBs detected earlier than GRB050904, the open circles express the others; the curve of $\varphi(\log f)$ is given by FIG. 16(*a*). we found that the maximum of the redshift of the GRB among the 71 GRBs is $z = 5.3$ (GRB050814); if we wish to forecast the value of the fluence of the GRB that redshift would exceed or close to 5.3 and will soon be detected, we are sure, according to FIG. 16(*b*), that the expected value should take $\varphi(\log f) = -5.4964 \pm 0.5036$ [i.e. $f = (1.0 \sim 10.2) \times 10^{-6}$ erg cm$^{-2}$]. We found that this expectation is quite exact; the *five* samples with the redshift $z > 4.5$ and detected later than GRB050814 indeed all tally with the prediction without exception.

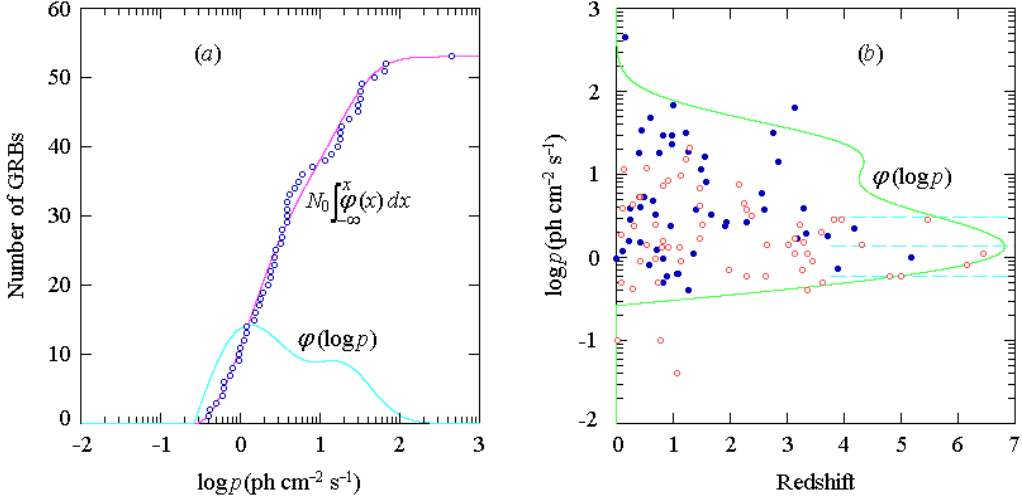

**FIG. 17.** Distribution function and forecast. (*a*) Distribution function $\varphi(\log p)$ for 53 GRBs detected earlier than GRB050904. (*b*) The *p-z* relation of 111 GRBs; where, the solid circles express the 53 GRBs detected earlier than GRB050904; the open circles express the others.

FIG. 17(*a*) showed the distribution function $\varphi(\log p)$ of the photon flux for the earliest 54 GRBs that detected earlier than GRB050904; FIG. 17(*a*) gives the following distribution function of $\varphi(\log p)$:

$$\varphi(x) = 0.306\exp[-4(x-1.32)^2] + c(x+0.58)\exp[-(x+0.58)^2]. \quad (9.2)$$

Where, $x = \log f$; $c = 0$ if $x < -0.58$, otherwise $c = 1.458$; the maximum of $\varphi(\log p)$ is at $\log p = 0.1311$.

FIG. 17(*b*) showed the log $p-z$ relation for whole 111 GRBs, in which the solid circles express the 54 GRBs detected earlier than GRB050904, the open circles express the others; the curve of $\varphi(\log p)$ is given by FIG. 17(*a*). we found that the maximum of the redshift of the GRB among the 54 GRBs is $z = 5.3$ (GRB050814); if we wish to forecast the value of the photon flux of the GRB that redshift would exceed or close to 5.0 and will soon be detected, we are sure, according to FIG. 17(*b*), that the expected value would take $\log p = 0.1311 \pm 0.3529$ (i.e. $p = 0.6 \sim 3.05$ ph cm$^{-2}$ s$^{-1}$). We found that this expectation is quite exact; the *eight* samples with the highest redshift and detected later than GRB050814 indeed all tally with the prediction without exception.

Above examples had in fact provided a solid verification for the independent physical quantity law.

Along with the growth of the sample size of GRBs and the change of the observed conditions, some changes of the distribution functions for the fluence and the photon flux, i.e. the log $f-z$ relation and the log $p-z$ relation, have occurred. We making the predictions must use the new results, i.e. use the FIG. 13(*b*) and FIG. 14(*b*). According to FIG. 13(*b*), we are sure that the expected value of the fluence $f$ of the GRB, which redshift would exceed or close to 4.5 and will soon be detected, would take $\log f = -5.6182 \pm 0.3818$ [i.e. $f = (1.0 \sim 5.8) \times 10^{-6}$ erg cm$^{-2}$]; and according to FIG. 14(*b*), we are sure that the expected value of the 1 s peak photon number flux $p$ of the GRB, which redshift would exceed or close to 4.5 and will soon be detected, would take $\log p = 0.157 \pm 0.379$ [i.e. $p = (0.6 \sim 3.4)$ ph cm$^{-2}$ s$^{-1}$]. The expected values of both the fluence and the photon flux all concentrate in a narrow region.

Using the existing distribution data of peak photon flux of 64 ms and 256 ms to the redshift, we can also forecast the value of peak photon flux of 64 ms and 256 ms for those GRBs which redshift would exceed or close to 4.5 and will soon be detected.

## 10. CONCLUSIONS AND DISCUSSIONS

The analyses above including the correlation analysis, the deductive reasoning and the variables separation test all deduce the same conclusion that the distance of GRBs is independent from its redshift. There is no any signature to show that the redshift can (or must) indicate the distance; if we compel the redshift to indicate the distance of GRBs, we then definitely obtain a series of wrong results that violates the basic principle of physics and entirely cannot tally with the observed facts.

Why the redshift of GRBs cannot indicate the distance but that of quasars does? There are two possibilities, perhaps it originates from the gravitational redshift of a neutron star, perhaps it come from the background galaxy, or part of GRBs is the former, others the latter. The redshift cannot indicate the distance only under these circumstances. However, if the redshift of GRBs originates from the gravitational redshift, the linear number-redshift relation of GRBs would be hardly understood. It has demonstrated above that the precise background galaxy theory can successfully explain all the distribution laws of GRBs, this suggests that the redshift of at least the most of GRBs would come from the background galaxy.

How distance the GRB would have? GRB030329 provided a reliable clue; the observation shows that the mean expansion velocity of its afterglow is $3\times10^4$ km s$^{-1}$ in the first 25 days (Hjorth et al. 2003a; Pian et al. 2006), the angular size of the afterglow is 0.07 mas at the 25th day after the burst (Taylor et al. 2004), we then obtain the distance of GRB030329 to be ~ 13 Mpc. Due to GRB030329 is the brightest GRB, the distance would be the closest one, other GRBs would have greater distance; according to the result given by the analysis for the data obtained by BATSE, see also §2 above, the distribution range of GRBs would be less than 350 Mpc. Due to the instruments threshold, farther GRBs are difficult to detect.

In a word, GRBs in fact come from the "shallow" rather than the "deep" place of the Universe.

**ACKNOWLEDGMENTS**

**Table 1 Summary of the observed data of GRBs**

| GRB | Redshift [a] | | Fluence (erg cm$^{-2}$) | Peak flux (ph cm$^{-2}$ s$^{-1}$) | Host [b] (R-mag) | Ref. [c] |
|---|---|---|---|---|---|---|
| 970228 | 0.695 | E | 6.45E-6 | 3.300 | 25.50 | 1 |
| 970508 | 0.8349* | EA | 3.96E-6 | 0.970 | 25.72 | 2, 3 |
| 970828 | 0.9579 | E | 9.60E-5 | — | 24.50 | 4 |
| 971214 | 3.418* | E | 1.25E-5 | 1.950 | 25.60 | 3, 5 |
| 980425 | 0.0085 | E | 4.01E-6 | 0.960 | — | 3, 6 |
| 980613 | 1.0964 | E | 1.71E-6 | 0.630 | 23.85 | 7 |
| 980703 | 0.9661* | EA | 6.22E-5 | 2.400 | 22.41 | 3, 8 |
| 990123 | 1.6004* | A | 4.87E-4 | 16.410 | 23.77 | 3, 9 |
| 990506 | 1.30658 | E | 2.11E-4 | 18.560 | 25.00 | 3, 10 |
| 990510 | 1.619* | A | 2.06E-5 | 8.170 | 28.50 | 3, 11 |
| 990705 | 0.8424 | E | 9.30E-5 | — | 22.20 | 12 |
| 990712 | 0.4331* | EA | 2.56E-5 | 4.100 | 21.91 | 13 |
| 991208 | 0.7063* | E | 1.00E-4 | — | 24.27 | 14 |
| 991216 | 1.02* | A | 2.51E-4 | 67.520 | 24.50 | 3, 15 |
| 000131 | 4.5 | A | 1.00E-5 | — | — | 16 |
| 000210 | 0.8463 | E | 6.10E-5 | 29.900 | 23.50 | 17 |
| 000301C | 2.0404* | A | 4.10E-6 | — | 28.00 | 18 |
| 000418 | 1.11854 | E | 1.61E-6 | 0.630 | 23.90 | 3, 19 |
| 000911 | 1.0585 | E | 2.30E-4 | — | 25.41 | 20 |
| 000926 | 2.0379* | A | 2.20E-5 | — | 25.00 | 21 |
| 001109 | 0.398 | E | 4.97E-6 | — | 20.70 | 22 |
| 010222 | 1.47755* | A | 9.20E-5 | — | — | 23 |
| 010921 | 0.451 | E | 1.84E-5 | 34.000 | 21.70 | 24, 25 |
| 011121 | 0.362* | E | 2.00E-5 | — | 21.20 | 26 |
| 011130 | 0.50* | A | 6.80E-7 | 5.400 | — | 25, 27 |
| 011211 | 2.142* | EA | 5.00E-6 | — | 25.00 | 28 |
| 020124 | 3.198* | A | 8.10E-6 | 64.000 | 29.30 | 25, 29 |
| 020405 | 0.6908* | EA | 3.00E-5 | — | — | 30 |
| 020531 | 1.00* | A | 1.23E-6 | 23.000 | — | 25, 31 |
| 020813 | 1.2545* | EA | 9.79E-5 | 32.000 | — | 25, 32 |
| 020819 | 0.41 | A | 8.80E-6 | 18.000 | 19.50 | 25, 33 |
| 020903 | 0.251* | E | 1.00E-7 | 2.800 | — | 25, 34 |
| 021004 | 2.3351* | EA | 2.60E-6 | 2.700 | 24.36 | 25, 35 |
| 021211 | 1.006 | E | 3.53E-6 | 30.000 | — | 25, 36 |
| 030226 | 1.98691* | A | 5.60E-6 | 2.700 | — | 25, 37 |
| 030227 | 1.39* | E | 7.50E-7 | 1.100 | — | 38 |
| 030323 | 3.3718* | A | 1.20E-6 | 3.900 | — | 25, 39 |
| 030328 | 1.5216* | A | 3.70E-5 | 11.600 | — | 25, 40 |
| 030329 | 0.16854* | EA | 1.63E-4 | 451.000 | 22.70 | 25, 41 |
| 030429 | 2.6564* | A | 1.50E-6 | 3.800 | — | 25, 42 |
| 030528 | 0.782 | E | 1.19E-5 | 17.900 | — | 25, 43 |
| 031203 | 0.1055 | E | 4.00E-7 | 1.200 | — | 44 |
| 040511 | 2.63 | E | 1.00E-5 | — | — | 45 |
| 040701 | 0.2146 | E | 4.50E-7 | — | — | 46 |
| 040827 | 0.9* | A | — | 0.600 | 23.10 | 47 |
| 040912 | 1.563 | E | 1.16E-6 | — | — | 48 |
| 040924 | 0.859 | E | 4.20E-6 | — | — | 49 |
| 041006 | 0.712* | EA | 1.20E-5 | — | — | 50 |
| 050126 | 1.29 | E | 1.10E-6 | 0.400 | — | 51 |
| 050223 | 0.5915 | E | 7.40E-7 | 0.800 | 21.60 | 52 |
| 050315 | 1.949* | A | 4.20E-6 | 2.400 | 23.90 | 53 |
| 050318 | 1.44* | A | 2.10E-6 | 3.800 | — | 54 |
| 050319 | 3.24* | A | 8.00E-7 | 1.700 | — | 55 |
| 050401 | 2.9* | A | 1.93E-5 | 14.000 | — | 56 |
| 050406 | 2.44 | | 9.00E-8 | 3.200 | — | 57 |
| 050408 | 1.2357* | EA | 3.30E-6 | — | — | 58 |
| 050416 | 0.6535 | E | 3.80E-7 | 4.800 | — | 59 |
| 050502 | 3.793* | A | 1.40E-6 | 1.800 | — | 60 |
| 050505 | 4.2748* | A | 4.10E-6 | 2.200 | — | 61 |
| 050509B | 0.2249* | A | 2.30E-8 | 1.570 | — | 62 |
| 050525 | 0.606* | EA | 7.84E-5 | 48.000 | — | 63 |
| 050603 | 2.821* | E | 3.41E-5 | 31.800 | — | 64 |
| 050709 | 0.16 | E | 1.40E-6 | — | 21.11 | 65 |
| 050724 | 0.257* | A | 6.30E-7 | 3.900 | — | 66 |
| 050730 | 3.96855* | A | 4.40E-6 | 0.740 | — | 67 |
| 050802 | 1.71* | A | 2.80E-6 | 3.300 | — | 68 |
| 050803 | 0.422* | E | 3.90E-6 | 1.500 | — | 69 |
| 050813 | 0.722* | A | 1.24E-7 | 1.220 | — | 70 |
| 050814 | 5.3 | | 2.17E-6 | 1.000 | — | 71 |
| 050820 | 2.6147* | A | 8.40E-6 | 6.000 | — | 72 |
| 050824 | 0.83* | EA | 2.30E-7 | 0.500 | — | 73 |
| 050826 | 0.297 | E | 4.30E-7 | 0.420 | 21.67 | 74 |
| 050904 | 6.29* | A | 5.40E-6 | 0.800 | — | 75 |
| 050908 | 3.3437* | A | 5.10E-7 | 0.700 | — | 76 |
| 050922C | 2.198* | A | 7.30E-6 | 7.500 | — | 77 |
| 051016B | 0.9364* | E | 1.70E-7 | 1.320 | — | 78 |
| 051022 | 0.807* | EA | 2.61E-4 | — | — | 79 |
| 051109 | 2.346* | A | 4.00E-6 | 3.700 | — | 80 |
| 051109B | 0.080 | A | 2.70E-7 | 0.500 | — | 81 |
| 051111 | 1.54948* | A | 8.40E-6 | 2.500 | — | 82 |
| 051221 | 0.5464* | E | 3.20E-6 | 12.100 | — | 83 |
| 051227 | 0.714* | EA | 2.30E-7 | 0.970 | — | 84 |
| 060108 | 2.03 | | 3.70E-7 | 0.700 | — | 85 |
| 060115 | 3.53* | A | 1.90E-6 | 0.900 | — | 86 |
| 060116 | 6.6 | | 2.60E-6 | 1.100 | — | 87 |
| 060123 | 1.099* | E | 3.00E-7 | 0.040 | — | 88 |
| 060124 | 2.297* | A | 1.43E-5 | 4.500 | — | 89 |
| 060206 | 4.04795* | A | 8.40E-7 | 2.800 | — | 90 |
| 060210 | 3.91* | A | 7.70E-6 | 2.800 | — | 91 |
| 060218 | 0.03345* | E | 6.80E-6 | 0.100 | — | 92 |
| 060223 | 4.41* | A | 6.80E-7 | 1.400 | — | 93 |
| 060418 | 1.4901* | A | 1.60E-5 | 6.700 | — | 94 |
| 060502 | 1.51* | A | 2.20E-6 | 1.700 | — | 95 |
| 060502B | 0.287 | A | 4.00E-8 | 4.400 | — | 96 |
| 060505 | 0.089* | EA | 6.20E-7 | 1.900 | — | 97 |
| 060510B | 4.9* | A | 4.20E-6 | 0.600 | — | 98 |
| 060512 | 0.4428 | E | 2.30E-7 | 0.900 | — | 99 |
| 060522 | 5.11* | A | 1.10E-6 | 0.600 | — | 100 |
| 060526 | 3.21* | A | 4.90E-7 | 1.700 | — | 101 |
| 060602 | 0.787 | E | 1.60E-6 | 0.100 | — | 102 |
| 060604 | 2.68* | A | 1.30E-7 | 0.600 | — | 103 |
| 060605 | 3.711* | A | 4.60E-7 | 0.500 | — | 104 |
| 060607 | 3.082* | A | 2.60E-6 | 1.400 | — | 105 |
| 060614 | 0.125* | E | 4.09E-5 | 11.600 | — | 106 |
| 060707 | 3.425* | A | 1.70E-6 | 1.100 | — | 107 |
| 060714 | 2.71* | A | 3.00E-6 | 1.400 | — | 108 |
| 060729 | 0.54* | A | 2.70E-6 | 1.400 | — | 109 |
| 060801 | 1.131 | E | 8.10E-8 | 1.300 | — | 110 |
| 060814 | 0.84 | E | 2.69E-5 | 7.400 | — | 111 |
| 060904B | 0.703* | A | 1.70E-6 | 2.500 | — | 112 |
| 060906 | 3.685* | A | 2.21E-6 | 2.000 | — | 113 |
| 060908 | 2.43* | A | 2.90E-6 | 3.200 | 25.12 | 114 |
| 060912 | 0.937 | E | 4.00E-6 | 8.500 | — | 115 |
| 060926 | 3.208* | A | 2.20E-7 | 1.100 | — | 116 |
| 060927 | 5.6* | A | 1.10E-6 | 2.800 | — | 117 |
| 061004 | 3.3* | A | 5.70E-7 | 2.500 | — | 118 |
| 061006 | 0.4377* | A | 3.57E-6 | 5.360 | — | 119 |
| 061007 | 1.261* | EA | 2.49E-4 | 15.300 | — | 120 |
| 061110 | 0.757* | E | 1.10E-6 | 0.500 | — | 121 |
| 061110B | 3.44* | A | 1.30E-6 | 0.400 | — | 122 |
| 061121 | 1.314* | A | 5.67E-5 | 21.100 | — | 123 |
| 061126 | 1.1588 | E | 2.00E-5 | 9.800 | — | 124 |
| 061201 | 0.111* | E | 5.33E-6 | 3.900 | — | 125 |
| 061210 | 0.41* | E | 2.02E-6 | 5.300 | — | 126 |
| 061217 | 0.827* | E | 4.60E-8 | 1.300 | — | 127 |
| 061222B | 3.355* | A | 2.20E-6 | 1.500 | — | 128 |
| 070110 | 2.352* | A | 1.60E-6 | 0.600 | — | 129 |
| 070125 | 1.547* | A | 1.74E-4 | — | — | 130 |
| 070208 | 1.165* | EA | 4.30E-7 | 0.900 | — | 131 |
| 070209 | 0.314 | E | 1.10E-8 | 2.400 | — | 132 |
| 070306 | 1.497* | E | 5.50E-6 | 4.200 | — | 133 |
| 070318 | 0.84* | A | 2.30E-6 | 1.600 | — | 134 |
| 070411 | 2.954* | A | 2.50E-6 | 1.000 | — | 135 |
| 070419 | 0.97* | A | 5.60E-7 | 0.028 | — | 136 |
| 070429B | 0.904 | E | 6.30E-8 | 1.800 | — | 137 |
| 070506 | 2.31* | A | 2.10E-7 | 1.000 | — | 138 |
| 070521 | 0.553* | E | 1.81E-5 | 6.700 | — | 139 |
| 070529 | 2.4996* | A | 2.60E-6 | 1.400 | — | 140 |
| 070611 | 2.04* | A | 3.90E-7 | 0.800 | — | 141 |
| 070612 | 0.617* | E | 1.10E-5 | 1.500 | — | 142 |
| 070714B | 0.92 | E | 3.70E-6 | 2.700 | — | 143 |
| 070721B | 3.626* | A | 2.10E-6 | 1.500 | — | 144 |
| 070724 | 0.457* | E | 3.00E-8 | 1.000 | — | 145 |
| 070802 | 2.45* | A | 2.80E-7 | 0.400 | — | 146 |
| 070810 | 2.17* | A | 6.90E-7 | 1.900 | — | 147 |
| 071003 | 1.100* | A | 5.32E-5 | 6.300 | — | 148 |
| 071010 | 0.98* | A | 2.00E-7 | 0.800 | — | 149 |
| 071010B | 0.947* | EA | 4.78E-6 | 7.700 | — | 150 |
| 071020 | 2.145* | A | 7.71E-6 | 8.400 | — | 151 |
| 071031 | 2.692* | A | 9.00E-7 | 0.500 | — | 152 |
| 071112C | 0.823* | EA | 3.00E-6 | 8.000 | — | 153 |
| 071117 | 1.331* | E | 5.84E-6 | 11.300 | — | 154 |
| 071122 | 1.14* | A | 5.80E-7 | 0.400 | — | 155 |
| 071227 | 0.384* | E | 1.60E-6 | 1.600 | — | 156 |

[a] Symbol "A" expresses that the redshift is determined by the absorption lines, "E" by the emission lines, "EA" by both absorption and emission lines; the asterisk (*) implies that the redshift obtained from the afterglow.

[b] Containing a part of $R_c$-magnitude and $V$-magnitude.

[c] References — (1) Djorgovski et al. 1999c; Amati et al. 2002; Palmer et al. 1997; Djorgovski et al. 1997; (2) Metzger et al. 1997; Bloom et al. 1998; (3) BATSE; (4) Djorgovski et al. 2001a; Piran et al. 2000; Djorgovski 2001c; (5) Odewahn et al. 1998; Kulkarni et al. 1998; (6) Tinney et al. 1998; (7) Djorgovski et al. 1999a; Woods et al. 1998; (8) Holland et al. 2001; (9) Djorgovski et al. 1999b; Halpern et al. 1999; (10) Bloom 2003; Holland et al. 2000; (11) Vreeswijk et al. 1999a; Bloom 2000; (12) Le Floc'h et al. 2002; Lazzati et al. 2001; Saracco et al. 2001; (13) Vreeswijk et al. 2001; Frontera 2001; Hjorth et al. 2000; (14) Castro-Tirado et al. 2001; Hurley & Cline 1999; (15) Vreeswijk et al. 1999b; Djorgovski et al. 1999d; (16) Andersen et al. 2001; Hurley et al. 2000 (17) Piro et al. 2002; Kippen 2000; (18) Jensen et al. 2001; Fruchter & Vreeswijk 2001; (19) Bloom et al. 2000; Metzger et al. 2000; (20) Price et al. 2002; Price et al. 2001b; (21) Castro 2003; Hurley 2000; Harrison et al. 2001; (22) Afanasiev et al. 2001; Guidorzi et al. 2003; Greiner et al. 2000; (23) Jha et al. 2001a; in't Zand et al. 2001; (24) Djorgovski et al. 2001b; 265; (25) Sakamoto et al. 2005a; (26) Garnavich et al. 2003; Brown et al. 2001; (27) Jha et al. 2001b; (28) Gladders et al. 2001; Frontera et al. 2002; Burud et al. 2001; (29) Hjorth et al. 2003b; Bloom et al. 2002; (30) Masetti et al. 2003; Hurley et al. 2002; (31) Kulkarni et al. 2002; (32) Fiore et al. 2002; (33) Jakobsson et al. 2005a; (34) Soderberg et al. 2004; (35) Møller et al. 2002; Fatkhullin et al. 2002; (36) Vreeswijk et al. 2003; (37) Shin et al. 2006; (38) Watson et al. 2003; Mereghetti et al. 2003; (39) Vreeswijk et al. 2004; (40) Maiorano et al. 2006; (41) Bloom et al. 2003; Fruchter et al. 2003; (42) Weidinger et al. 2003; Hurley et al, 2003; (43) Rau et al. 2005; (44) Prochaska et al., 2004; Mereghetti & Gotz 2003; (45) Berger et al. 2004; Dullighan et al. 2002; (46) Kelson et al. 2004; Barraud et al. 2004;

2006b; (129) Jaunsen et al. 2007a; Cummings et al. 2007a; (130) Fox et al. 2007; Golenetskii et al. 2007a; (131) Cucchiara et al. 2007a; Markwardt et al. 2007; (132) Berger & Fox 2007; Sakamoto et al. 2007a; (133) Jaunsen et al. 2007b; Barthelmy et al. 2007a; (134) Chen et al. 2007; Cummings et al. 2007b; (135) Jakobsson et al. 2007a; Markwardt 2007a; (136) Cenko et al. 2007a; Stamatikos 2007a; (137) Perley et al. 2007c; Tueller 2007a; (138) Christina et al. 2007; Barbier 2007a; (139) Hattori et al. 2007; Golenetskii et al. 2007b; Palmer 2007; (140) Berger et al. 2007c; Parsons 2007a; (141) Thoene et al. 2007a; Barbier et al. 2007b; (142) Cenko et al. 2007b; Barthelmy et al. 2007b; (143) Graham et al. 2007; Sakamoto et al. 2007b; Barbier et al. 2007c; (144) Malesani et al. 2007; Palmer et al. 2007; (145) Cucchiara et al. 2007b; Parsons 2007b; (146) Prochaska et al. 2007a; Cummings 2007; (147) Thoene et al. 2007c; Markwardt 2007b; (148) Perley et al. 2007b; Golenetskii et al. 2007c; Ukwatta 2007; (149) Prochaska et al. 2007b; Krimm 2007a; (150) Stern et al. 2007; Golenetskii et al. 2007d; Markwardt 2007c; (151) Jakobsson et al. 2007b; Golenetskii et al. 2007e; Tueller 2007b; (152) Ledoux et al. 2007; Stamatikos 2007b; (153) Jakobsson et al. 2007d; Krimm 2007b; (154) Jakobsson et al. 2007e; Golenetskii 2007; Krimm 2007c; (155) Cucchiara et al. 2007c; Sakamoto 2007; (156) Berger et al. 2007b; Golenetskii et al. 2007f; Sato et al. 2007.